\documentclass[apj]{emulateapj}

\usepackage{amsmath,amstext,amssymb}
\usepackage{graphicx}
\usepackage{hyperref}
\usepackage[figure,figure*]{hypcap}
\usepackage{enumerate}
\usepackage{multirow}
\usepackage{amsfonts}
\usepackage{amssymb} 
\usepackage{bm}
\usepackage{natbib}
\usepackage{courier}
\usepackage{xcolor} 

\newcommand{\del}[1]{}

\newcommand{\comm}[1]{}

\shorttitle{The \textit{CHiPS} Survey: Complete sample of extreme BCG clusters}
\shortauthors{Somboonpanyakul et al.}

\altaffiltext{\MIT}{Kavli Institute for Astrophysics and Space Research, Massachusetts Institute of Technology, 77 Massachusetts Avenue, Cambridge, MA 02139}
\altaffiltext{\Princeton}{Department of Astrophysical Sciences, Princeton University, Princeton, NJ 08544, USA}
\altaffiltext{\INAF}{INAF, Osservatorio di Astrofisica e Scienza dello Spazio, via Piero Gobetti 93/3, 40129 Bologna, Italy}
\altaffiltext{\LSST}{Vera C. Rubin Observatory Project Office, 950 N. Cherry Ave, Tucson, AZ 85719, USA}
\altaffiltext{\CfA}{Center for Astrophysics | Harvard \& Smithsonian, 60 Garden Street, Cambridge, MA 02138, USA}

\def\MIT{1}
\def\Princeton{2}
\def\INAF{3}
\def\LSST{4}
\def\CfA{5}

\begin{document}

\title{The Clusters Hiding in Plain Sight (\textit{CH\MakeLowercase{i}PS}) survey: Complete sample of extreme BCG clusters}

\author{
Taweewat Somboonpanyakul\altaffilmark{\MIT},
Michael McDonald\altaffilmark{\MIT},
Massimo Gaspari\altaffilmark{\Princeton,\INAF}, 
Brian Stalder\altaffilmark{\LSST,\CfA}, and
Antony A. Stark\altaffilmark{\CfA}
}

\begin{abstract}
	We present optical follow-up observations for candidate clusters in the Clusters Hiding in Plain Sight (\textit{CHiPS}) survey, which is designed to find new galaxy clusters with extreme central galaxies that were misidentified as bright isolated sources in the ROSAT All-Sky Survey catalog. We identify 11 cluster candidates around X-ray, radio, and mid-IR bright sources, including six well-known clusters, two false associations of foreground and background clusters, and three new candidates which are observed further with \textit{Chandra}. Of the three new candidates, we confirm two newly discovered galaxy clusters: CHIPS1356-3421 and CHIPS1911+4455. Both clusters are luminous enough to be detected in the ROSAT All Sky-Survey data if not because of their bright central cores. CHIPS1911+4455 is similar in many ways to the Phoenix cluster, but with a highly-disturbed X-ray morphology on large scales. We find the occurrence rate for clusters that would appear to be X-ray bright point sources in the ROSAT All-Sky Survey (and any surveys with similar angular resolution) to be $2\pm1\%$, and the occurrence rate of clusters with runaway cooling in their cores to be $<\!1\%$, consistent with predictions of Chaotic Cold Accretion. With the number of new groups and clusters predicted to be found with \textit{eROSITA}, the population of clusters that appear to be point sources (due to a central QSO or a dense cool core) could be around 2000. Finally, this survey demonstrates that the Phoenix cluster is likely the strongest cool core at $z<0.7$ -- anything more extreme would have been found in this survey.
\end{abstract}


\section{Introduction}
Clusters of galaxies are the largest and most massive gravitationally bound objects in the universe, with masses of roughly $10^{14}-10^{15}M_\odot$~\citep{2005Voitb} and extended on scales of several Mpc. On this scale, the density field remains in the linear regime of density perturbation~\citep{1991Henry}. This means that the number density of clusters can be predicted based on first principles (see~\citet{2008Tinker} for the most recent calibration of halo mass function). This number density depends strongly on several cosmological parameters, including $\Omega_m$ (the density of total matter compare) and $\sigma_8$ (the amount of fluctuation in matter density)~\citep{2009Vikhlininb}. This forms the basis of cluster cosmology.

Since the end of the \textit{Planck} Satellite's mission~\citep{2018Planck-VI}, we are now living in the era of precision cosmology where cosmological parameters of the universe are routinely measured with percent-level uncertainty. To improve the precision of cluster cosmology, various groups have been trying to increase the number of known galaxy clusters, by searching for overdensities of red galaxies in optical or near-infrared~\citep{2000Gladders,2014Rykoff,2019Gonzalez}, extended extragalactic emission in X-ray~\citep{2000Ebeling,2001Ebeling,2004Bohringer}, or via Sunyaev-Zel'dovich (SZ) effect~\citep{1972Sunyaev,2015Bleem,2019Bleem,2016Planck-catalog,2020Hilton} in millimeter/sub-millimeter surveys. Each technique has its own unique benefits and challenges. With the invention of wide field optical telescopes, performing optical surveys to find overdensities of galaxies is relatively cheap, although optical surveys are strongly affected by projection effects. For SZ surveys, we are capable of detecting galaxy clusters up to relatively high redshift since the SZ signature is redshift independent. On the other hand, the SZ signature depends strongly on mass, restricting current-generation surveys to only the most massive clusters~\citep{2002Carlstrom,2005Motl}. Lastly, X-ray surveys have been one of the most popular techniques to discover galaxy clusters since the launch of the ROSAT X-ray satellite (e.g., the REFLEX survey~\citep{2004Bohringer}). Even though X-ray surveys can only produce flux-limited samples of galaxy clusters, cosmologists can take that into account in their selection function when they estimate cosmological parameters~\citep{2008Allen,2009Vikhlininb,2015Mantz}. However, with the continuous improvement in optical cluster finders, many of these SZ and X-ray cluster catalogs are now confirmed by the optical data, such as the recent works with SZ~\citep{Bleem2020} and X-ray catalogs~\citep{Klein2019}.

With the recent SZ discovery of the Phoenix cluster~\citep{2011Williamson,2012McDonald,2015McDonald}, the most X-ray luminous galaxy cluster known, at $z=0.6$, we have started to question our understanding of the X-ray-survey selection function. The Phoenix cluster was detected in several previous X-ray surveys, but was misidentified as a bright point source based on its extremely bright active galactic nucleus (AGN) and cool core in the center of the cluster. With most X-ray surveys identifying objects as either a point-like or an extended source, a galaxy cluster with a bright point source in the center could be misidentified as simply a point source. The next logical step is to ask how many of these galaxy clusters we have missed in the previous surveys, and how this translates to a correction for the selection function.

Another benefit of finding galaxy clusters hosting bright X-ray point sources is to study the cooling flow problem, which is the apparent disagreement between the X-ray luminosity (cooling rate) of a cluster and the observed star formation rate, the latter which is typically suppressed by a factor of $\sim\!100$. 
The best candidate for explaining the inconsistency is AGN feedback from the central galaxy~\citep{2006Bower,2006Croton,2008Bower}. There are two main modes of AGN feedback: the kinetic mode, driven mostly by jets, and the radiative mode, driven by the accretion of the AGN~\citep{2012Fabian,2012McNamara,2017Harrison,2020Gaspari}. With very few known galaxy clusters with extremely bright quasars, such as H1821+643~\citep{2010Russell}, 3C 186~\citep{2005Siemiginowska,2010Siemiginowska}, 3C 254~\citep{2003Crawford,2018Yang}, IRAS09104+4109~\citep{2012OSullivan}, and the Phoenix cluster~\citep{2012McDonald}, a larger number of such objects are required to fully understand the role of radiative-mode feedback in the evolution and formation of galaxy clusters. For example, the Chaotic Cold Accretion (CCA) model predicts a tight co-evolution between the central supermassive black hole (SMBH) and the host cluster halo~\citep[via the cooling rate or $L_{\rm x}$;][]{2019Gaspari}, with flickering quasar-like peaks reached only a few percent of times~\citep{2017Gaspari}.

In an attempt to find more galaxy clusters hosting bright central point sources, we started the Clusters Hiding in Plain Sight (\textit{CHiPS}) survey. The details and the first discovery from the survey is published in~\citet{2018Somboonpanyakul}. In this paper, we focus on a new optical cluster finding algorithm, developed specifically for the \textit{CHiPS} survey, to look for cluster candidates after optically imaging all of the X-ray point sources with bright radio and mid-IR from the first part of the project. These candidates may have been misidentified in previous all-sky surveys due to their central galaxies' brightness. After performing the cluster finding algorithm, we present a list of newly-discovered galaxy cluster candidates along with their expected redshift and richness.

The overview of the \textit{CHiPS} survey, the optical data used in the follow-up campaign, and its methodology are described in Section~\ref{sec::survey}. In Section~\ref{sec::pisco}, we present details of the data reduction and analysis for recently-obtained optical data from the Magellan telescope. Our cluster finding algorithm is described in Section~\ref{sec::algorithm} while the X-ray data reduction is presented in Section~\ref{sec::xray}. We discuss the results and the implications of these findings in Section~\ref{sec::result} and~\ref{sec::diss}. Lastly, we summarize the paper in Section~\ref{sec::summary}. We assume $H_0 = 70\,\rm{km\,s^{-1}\,Mpc^{-1}}$, $\Omega_{\rm m}= 0.3$ and $\Omega_{\rm \lambda} = 0.7$. All errors are $1\sigma$ unless noted otherwise.

\section{the \textit{CHiPS} Survey}~\label{sec::survey}
The \textit{CHiPS} survey is designed to identify new centrally concentrated galaxy clusters and clusters hosting extreme central galaxies (starbursts and/or AGNs) within the redshift range 0.1--0.7. The first part of the survey consists of identifying candidates by combining several all-sky survey catalogs to look for bright objects at multiple wavelengths. The second part of the survey, which is the focus of this paper, addresses mainly our optical follow-up program to determine the best cluster candidates by searching for an overdensity of galaxies at a given redshift centered on the location of the X-ray sources. The last part, which is also included in this paper, is to obtain \textit{Chandra} data for these candidates in order to confirm the existence of these new clusters and characterize their properties, such as the gas temperature, the total mass, and the gas fraction. 

\subsection{Target Selection}\label{sec::target}
Our \textit{CHiPS} target selection is described in detailed in our previous publication~\citep{2018Somboonpanyakul}; here we outline the main steps. 

To select systems similar to the Phoenix cluster, we require sources to be bright in X-ray, mid-IR and radio, relative to near-IR. The normalization to near-IR is to prevent very nearby (e.g., galactic) low-luminosity sources from overwhelming the sample. Starting with X-ray point source catalogs from the \textit{ROSAT} All-Sky Survey Bright Source Catalog and Faint Source Catalog~\citep[RASS-BSC and RASS-FSC;][]{1999Voges}, we cross-correlate with radio from NVSS~\citep{1998Condon} or SUMSS~\citep{2003Mauch}, mid-IR with WISE~\citep{2010Wright}, and near-IR with 2MASS~\citep{2006Skrutskie}. This combination leads to two types of astrophysical sources: radio-loud type II QSOs and galaxy clusters with an active core (a starburst and/or AGN-hosting BCG). This approach is similar to two other surveys from~\citet{2017Green} and~\citet{2020Donahue}. The main difference is that~\citet{2020Donahue} focus on previously-known optically-selected BCGs from the GMBCG catalog~\citep{2010Hao} and~\citet{2017Green} started with spectroscopically confirmed AGN in the ROSAT catalog. We begin our search with a complete ROSAT point source catalog and combine with all archival data from near-IR, mid-IR to radio.

\begin{figure*}[!ht]
	\begin{center}
		\includegraphics[width=1.96\columnwidth]{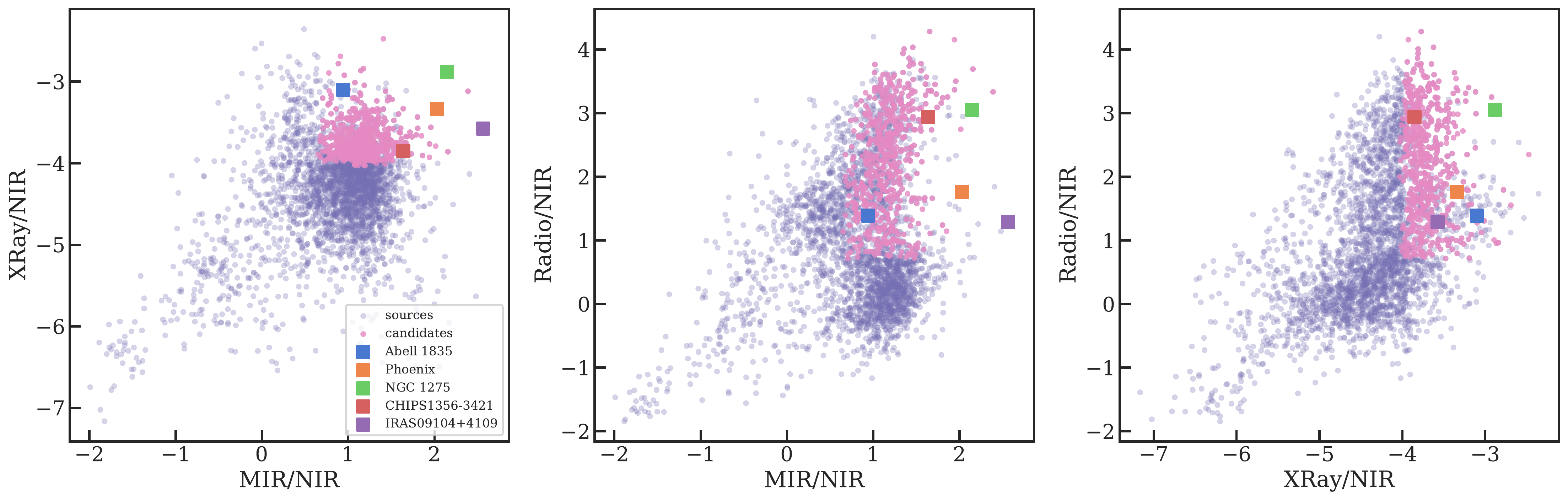}
		\caption{The three panels show color-color diagrams for objects that are detected in all four all-sky surveys (3,450 objects). The axes are the logarithm of the ratio of the X-Ray, mid-IR (MIR) or radio flux to the near-IR (NIR) flux. Points colored in pink satisfy our three color cuts. The Phoenix, Perseus (NGC 1275), Abell 1835, and IRAS09104+4109 clusters, which host extreme BCGs, are shown with orange, green, blue and purple squares, respectively while CHIPS1356-3421 is shown with a red square.}
		\label{fig::color-cut}
	\end{center}
\end{figure*}

In addition, we apply color cuts in order to select only the most extreme objects in X-ray, mid-IR, and radio, as demonstrated in Fig.~\ref{fig::color-cut}. The cuts are chosen to capture the expected range of color for a Phoenix-like object at an unknown redshift between 0.1 and 0.7. The NASA/IPAC Extragalactic Database (NED)\footnote{https://ned.ipac.caltech.edu} was used to reject foreground ($z < 0.1$) and background ($z > 0.7$) objects. Candidates with $z < 0.1$ are close enough to be detected with past instruments even with a bright central point source. Most of these clusters were first detected by eye in various optical catalogs, including the well-known Abell and Zwicky catalogs~\citep{Abell1989,Zwicky1961}, meaning that we do not expect any misclassifications. On the other hand, clusters at $z > 0.7$ are exceedingly rare in the ROSAT data -- not because of a bias in their selection, but because they are simply too faint. We also remove objects which have galactic latitude less than $\pm15^{\circ}$ because foreground stars and extinction from the Milky Way will obscure any clusters. After the removal, we are left with 470 objects to perform the optical follow-up, which is presented in the upcoming section. We note that by requiring mid-IR and radio detection, we emphasize the detection of Phoenix-like clusters, at the expense of removing from the sample some BCGs with central AGN that are not radio-loud or mid-IR-bright, such as unobscured, radio-quiet AGN. This means that the \textit{CHiPS} survey will only place a lower limit of the fraction of clusters missed by the previous X-ray surveys.

Further information about our target selection and the first galaxy cluster discovered from this survey, the galaxy cluster surrounding PKS1353-341, are presented in~\citet{2018Somboonpanyakul}. In the next section, we describe the data used for our optical follow-up of these 470 candidates.

\subsection{Optical Follow-up Observations} \label{sec::optical}
\begin{figure}[!ht]
	\begin{center}
		\includegraphics[width=0.95\columnwidth]{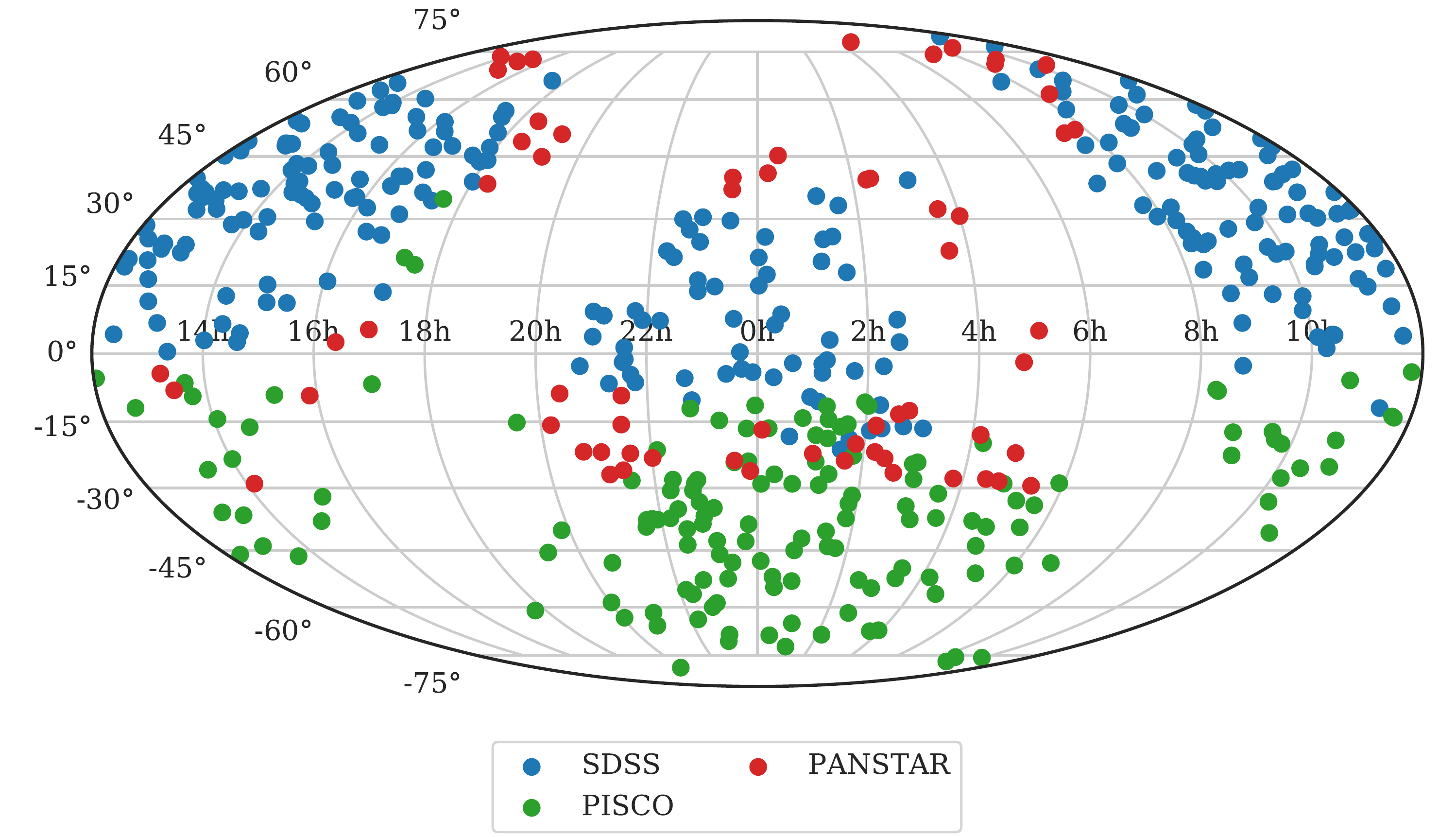}
		\caption{Plot of all 470 target candidates for the \textit{CHiPS} survey in the sky. The blue dots represented candidates followed-up with SDSS. The red dots are candidates followed-up with Pan-STARRS, and the green dots are candidates from PISCO observations. The gaps at RA = 18h-20h and 5h-7h corresponds to the Milky Way which prevents us from finding new cluster candidates around that region.}
		\label{fig::allsky}
	\end{center}
\end{figure}

The optical follow-up program is separated into two parts based on the declination of the targets. Most objects with positive declination are followed up with the Sloan Digital Sky Survey (SDSS) because of its nearly complete coverage in the Northern Sky. Whereas, objects with negative declination are observed with either the first data release of the Panoramic Survey Telescope and Rapid Response System~\citep[Pan-STARRS1;][]{2016Chambers} with sky coverage of declination greater than $-30^{\circ}$ or additional pointed observations using the Parallel Imager for Southern Cosmological Observations~\citep[PISCO;][]{2014Stalder} on the 6.5m Magellan Telescope at Las Campanas Observatory, Chile. Specifically, 256 out of our 470 candidates were observed with SDSS, 64 candidates were observed with Pan-STARRS1, and the remaining 150 candidates were individually observed with PISCO on the Magellan telescope. We note that data from the Dark Energy Survey~\citep[DES;][]{2016DES} was unavailable at the onset of the project. Fig.~\ref{fig::allsky} shows the position of all target candidates in the sky, separated by the telescope used for the follow-up.

\subsubsection{Sloan Digital Sky Survey (SDSS)}
The Sloan Digital Sky Survey (SDSS) is a multi-spectral imaging and spectroscopic redshift survey using a 2.5-m optical telescope at Apache Point Observatory in New Mexico~\citep{2006Gunn}. We utilized Data release 14 (DR14), released in 2017, which is the second data release for SDSS-IV~\citep{2018Abolfathi}. We retrieved the photometric data in $u$, $g$, $r$, $i$, and $z$ bands by querying objects within a radius of 5 arcmin from the X-ray position, using the function \texttt{fGetNearestObjEq} with the Casjob server\footnote{https://skyserver.sdss.org/CasJobs/}. In Section~\ref{sec::algorithm}, we apply a more stringent cut during the cluster finding algorithm. We obtained the SDSS model magnitude (\texttt{modelMag}) which, as explained in the SDSS support documentation\footnote{\label{ft::sdss}https://www.sdss.org/dr12/algorithms/magnitudes/}, gives the most unbiased estimates of galaxy colors. To convert SDSS magnitude to flux units, we use the SDSS asinh magnitude formula\textsuperscript{\ref{ft::sdss}}, which is also described in~\citep{1999Lupton}.

For star/galaxy classification, ``\texttt{type}'' parameters, provided by SDSS, were used to select only galaxies (\texttt{type} = 3). The classification is based on the difference between cmodel and PSF magnitude. Specifically, an object is classified as extended when $\rm{Mag_{PSF}}-\rm{Mag_{cmodel}} > 0.145$. In addition, we downloaded photometric redshifts ($z_{\rm{sdss}}$) for photometric redshift estimate verification in Section~\ref{sec::algorithm}.

\subsubsection{Panoramic Survey Telescope and Rapid Response System (Pan-STARRS)}
Pan-STARRS is a system for wide-field astronomical imaging in the optical $g$, $r$, $i$, $z$, and $y$ bands, located at Haleakala Observatory, Hawaii. The survey used a 1.8-m telescope, with an imaging resolution of $0.25^{\prime\prime}/\rm{pixel}$ from its 1.4 Gigapixel camera. Pan-STARRS1 (PS1), the basis for Data release 1 (DR1), covers three quarters of the sky ($3\pi$ survey) north of a declination of $-30^\circ$. 

\begin{figure}[!ht]
	\begin{center}
		\includegraphics[width=0.95\columnwidth]{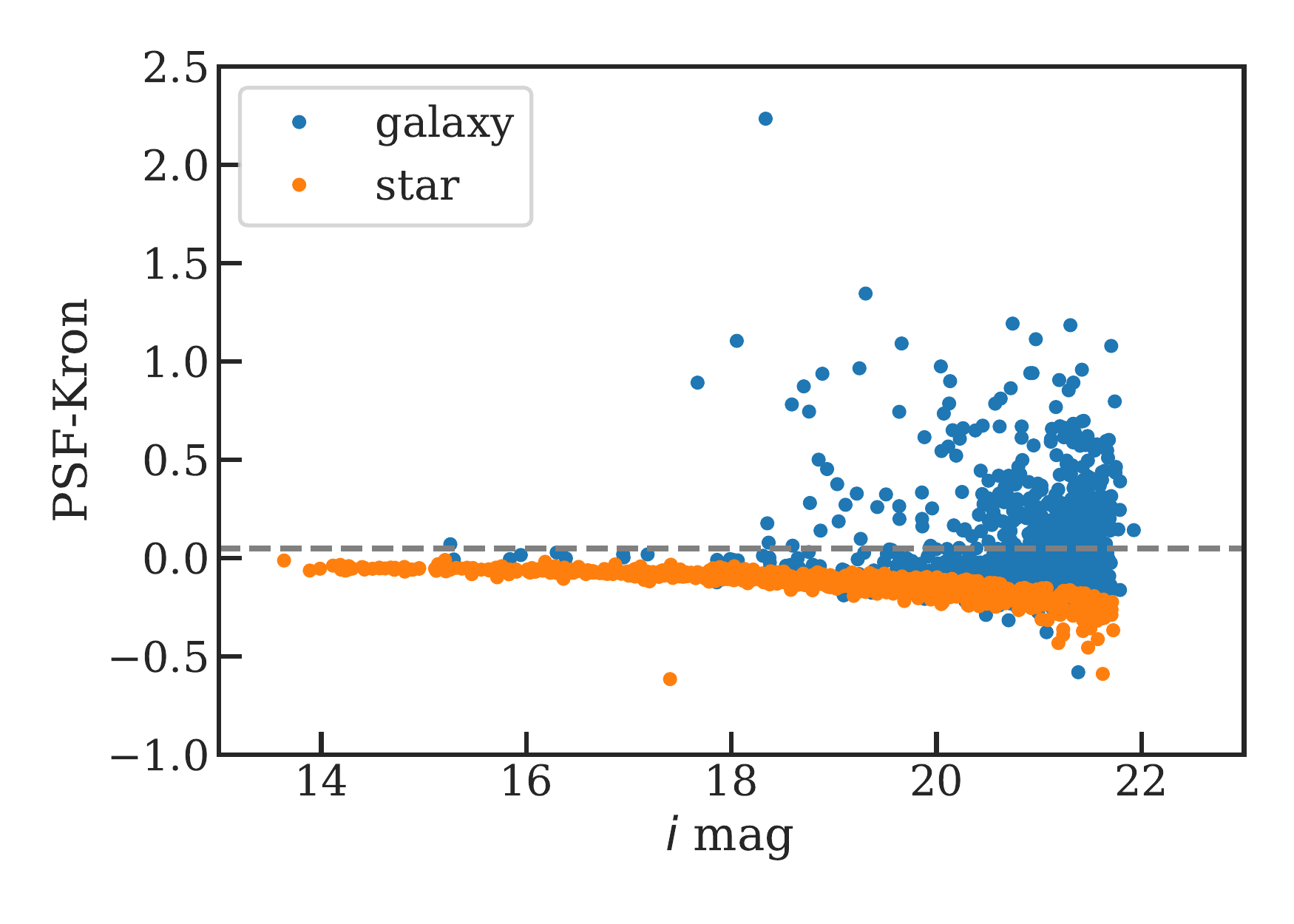}
		\caption{This figure demonstrates how we separate stars and galaxies in the Pan-STARRS sample catalog using the ratio of PSF and Kron magnitudes. The orange color indicates stars, which pass both the $r$ and $i$ band cut while the blue color marks galaxies.}
		\label{fig::s/g_panstar}
	\end{center}
\end{figure}

Star/galaxy separation of PS1 is similar to that of SDSS. Specifically, the difference between model and PSF magnitude is measured to identify extended objects. However, instead of applying a simple straight line as a cut (e.g., $\rm{Mag_{PSF}}-\rm{Mag_{Kron}} > 0.05$ where $\rm{Mag_{Kron}}$ is Kron magnitudes as the representation for model magnitude), an exponential model is used to fit the bright part of Fig.~\ref{fig::s/g_panstar} and then extrapolated to fainter objects, similar to \cite{2016Chambers}. Further details about this technique can be found in \cite{2014Farrow}.
As shown in Fig.~\ref{fig::s/g_panstar}, the star-galaxy separation is not a horizontal cut, but an exponential curve which takes into account our inability to distinguish between stars and galaxies at the fainter end. We require the cut to be satisfied for both $r$ and $i$ bands. Even though this star-galaxy-separation criterion could identify more objects as galaxies, this should not create a large bias for our cluster-finding algorithm. We chose Kron magnitudes as the Pan-STARRS magnitudes for our algorithm since they capture more light from the extended parts of galaxies, compared to PSF magnitudes.

\subsubsection{Magellan Telescope with PISCO}
Without a more robust all-sky survey in the southern sky similar to SDSS and Pan-STARRS in the north, we perform 150 individual follow-up observations for targets in the southern sky with the 6.5-m Magellan telescope. PISCO, a multi-band photometer, is used to speed up our observations because of our large number of candidates. With the ability to produce $g$, $r$, $i$, and $z$ band images simultaneously, our effective efficiency in observing these candidates increases by a factor of $\sim\!3$ ~\citep[including optical losses;][]{2014Stalder}. All candidates were acquired with PISCO during 9 nights splitting over 3 observing runs between 2017 January to 2017 December. We observed most objects with 5-minute total exposure with two 2.5-minute exposures for dithering. To analyze the PISCO data, we have created a data processing pipeline. Further details regarding data reduction and star/galaxy separation for the \textit{CHiPS} survey are presented in the next section. 

\subsection{X-ray Follow-up Observations}
To confirm the existence of a galaxy cluster, we require the detection of extended X-ray emission, indicating an extremely hot intracluster medium (ICM), reflecting the deep potential well of the cluster. The \textit{Chandra} X-ray Observatory is best suited for the task, given that our targets may have bright central point sources. With an angular resolution of 0.5$^{\prime\prime}$, \textit{Chandra} has the capability to distinguish X-ray point sources (e.g., AGN) from the extended emission of the ICM. We observed a total of 5 additional candidates from the \textit{CHiPS} survey, apart from the initial sample of 4 candidates for the pilot study~\citep{2018Somboonpanyakul}. More details about the reduction process for the X-ray data can be found in Section~\ref{sec::xray}.

\section{PISCO Observations and Data Processing} \label{sec::pisco}
In this section, we describe the data reduction process for the PISCO data. Since we obtain raw images from the PISCO instrument on the Magellan telescope, we developed a complete reduction pipeline to convert these images to photometric catalogs for all galaxies in the field, which are then used as an input for our cluster finding algorithm. In contrast, SDSS and Pan-STARRS are wide-field surveys with available photometric catalogs and, as such, do not require any further data processing. 

\subsection{PISCO Image Reduction} \label{sec::img_reduc}
PISCO is a photometer that produces $g$, $r$, $i$, and $z$ band images simultaneously~\citep{2014Stalder}. The camera is composed of four 3k$\times$4k charge-coupled devices (CCDs), one for each of the four focal planes, with an un-binned scale of $0.109^{\prime\prime}$ per pixel, resulting in a $5^{\prime}\times9^{\prime}$ field of view. Each CCD is read out with two amplifiers.

For each image, the data reduction process consists of several steps, as follows. First, the median of all bias frames for each night is subtracted from both the median of all flat frames and the science frames. We do not attempt to quantify and remove the dark current, as it is negligible in these devices. The ratio between the two subtracted frames (flat and science) is the flat-fielded image. The two flat-fielded images (one from each amplifier) of a CCD are stitched together to create a complete image for each band ($g$, $r$, $i$, and $z$) and each exposure. L.A. Cosmic is run on each image for robust cosmic rays detection and removal~\citep{2001vanDokkum}. Astrometric calibration is carried out via Astrometry.net\footnote{http://astrometry.net}, which is used to find the absolute pointing, plate scale, orientation, and additional distortions in each image~\citep{2010Lang}.

\begin{figure}[!ht]
\begin{center}
	\includegraphics[width=0.95\columnwidth]{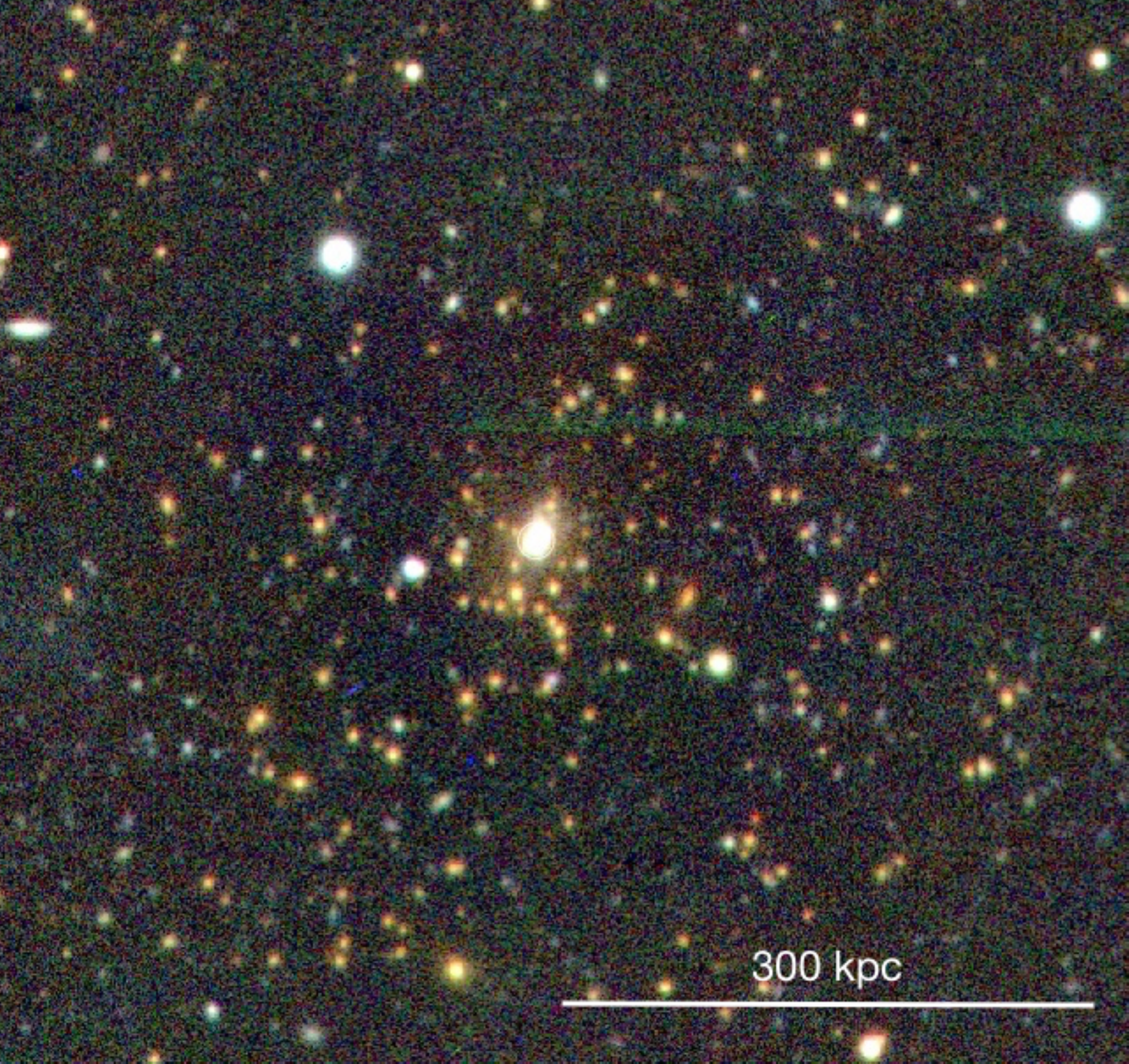}
	\caption{Three-color ($g$, $r$, $i$) image of the Phoenix cluster from PISCO. The image shows several red galaxies centrally located in the field, which is the signature of a galaxy cluster. The extremely bright point source in the center is a reason why the Phoenix cluster had been missed from previous X-ray surveys.}
	\label{fig::rgb_images}
\end{center}
\end{figure}

Multiple exposures need to be coadded to create a final, stacked image in each band. First, an initial source detection is run on all science images using \texttt{SExtractor}~\citep{1996Bertin}. Objects which are corrupted or truncated are removed from the lists by requiring the \texttt{FLAGS} parameter to be less than 5. Next, \texttt{SCAMP}~\citep{2002Bertin} is run over all of the images simultaneously to improve the astrometric solutions, previously obtained from Astrometry.net. The reference catalog we used for the astrometry is linked to the Two Micron All Sky Survey (2MASS) catalog~\citep{2006Skrutskie}. The individual images of each band are then resampled and coadded via \texttt{SWarp}~\citep{2006Bertin}. An example of the final processed image is shown in Fig.~\ref{fig::rgb_images}.

\subsection{Seeing Estimation and PSF Models} \label{sec::seeing}
\begin{figure}[!ht]
\begin{center}
\includegraphics[width=0.95\columnwidth]{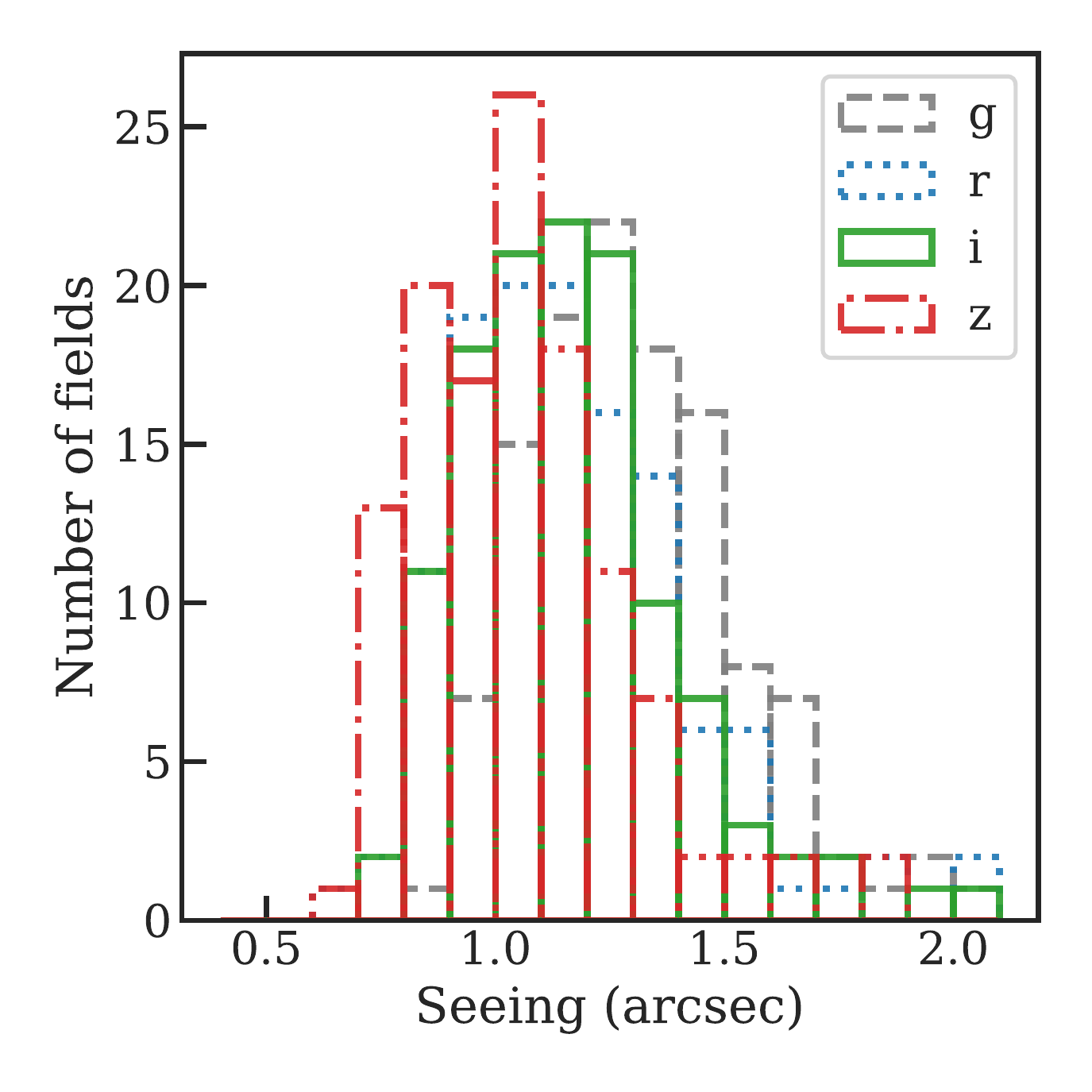}
\caption{Seeing distribution for the 262 fields observed with PISCO. The g band is traced by black dashed lines, the r band is traced by blue dotted lines, the i band is traced by green solid lines, and the z band is traced by red dot-dashed lines. The median values of the seeing in g, r, i, z bands are $1.^{\prime\prime}28$, $1.^{\prime\prime}15$, $1.^{\prime\prime}14$, $1.^{\prime\prime}03$ respectively.}
\label{fig::seeing}
\end{center}
\end{figure}

Even though each image already has an estimated seeing from the on-site seeing monitor, a more precise value is required for source extraction. We achieve this by fitting the Point Spread Function (PSF) models to every object in the field and picking the most common PSF to represent the seeing of that particular field. Specifically, we create $45\times45$-pixel small sub-image (``vignettes") for each detected object by using \texttt{SExtractor}. These small vignettes are fit using the 2D-Moffat model, available in the Astropy model packages~\citep{2018Astropy}. The Moffat model is a probability distribution that more accurately represents PSFs with broader wings than a simple Gaussian. We quote seeing measurements as the full width at half maximum (FWHM) of the best-fit model. Fig.~\ref{fig::seeing} shows the seeing distribution for objected observed with the PISCO camera. The median seeing in $g$, $r$, $i$, and $z$ bands are $1.^{\prime\prime}28$, $1.^{\prime\prime}15$, $1.^{\prime\prime}14$, and $1.^{\prime\prime}03$ respectively, meaning that the PSF tends to be broader for bluer bands, as expected. For one 2-night run on PISCO, the $r$-band data had a slightly worse (20\% larger) PSF due to alignment issues within the instrument. Given that this enlarged PSF is still smaller than that of the SDSS or Pan-STARRS data, we do not expect this to limit our analysis. The seeing distributions are not symmetric, but highly skewed toward higher seeing, representing a variation in the weather at the time of observation.

Apart from an accurate seeing estimate, the PSF model is also required for \texttt{SExtractor} to measure $\rm{MAG\_PSF}$. We use \texttt{PSFEx} to extract the PSF models from FITS images~\citep{2011Bertin}, setting all parameters to default. To get a good model for the PSF, we need to only select well-behaved point sources (stars) as our model. We achieve this by selecting sources which are not at the edges of the CCD, not elongated, and have an effective flux radius within $2\sigma$ of the mean for all sources in the field.

\subsection{Source Extraction} \label{sec::sextractor}
To measure an accurate color for each object, we extracted photometry via the dual-image mode of \texttt{SExtractor}, which uses the same pixel location for all photometric bands. The seeing estimates and the PSF models are used in this step with other input parameters described in Table~\ref{table::sextractor}. Next, we extract $griz$ $\rm{MAG\_AUTO}$, $\rm{MAG\_APER}$, and $\rm{MAG\_PSF}$ at the location of detected sources from the $i$-band image. 

\begin{deluxetable}{cc}
\centering 
\tabletypesize{\footnotesize} 
\tablecaption{\texttt{SExtractor} Source Detection Input Parameters\label{table::sextractor}}
\tablecolumns{15} 
\tablehead{ \colhead{Parameter}& \colhead{Value}} 
\startdata \vspace{0.0cm}
$\rm{DETECT\_MINAREA}$ & $1.1\pi\times(i\,\rm{band\,seeing}^2)$ \\
$\rm{DETECT\_THRESH}$ & 1.2\\
$\rm{GAIN}$ & 0.25 \\
$\rm{PIXEL\_SCALE}$\tablenotemark{a} & 0.12 or 0.22 \\
$\rm{SATUR\_LEVEL}$ & 61,000 \\
\enddata \vspace{-0.1cm}
\tablenotetext{a}{Depending on whether the data is binned.}
\end{deluxetable}

\subsection{Star-Galaxy Seperation} \label{sec::star_gal} 
One of the most important steps for the reduction pipeline is to separate sources into stars and galaxies. While $\rm{CLASS\_STAR}$\footnote{$\rm{https://sextractor.readthedocs.io/en/latest/ClassStar.html}$} is often used for this purpose, upon our close investigation we found non-negligible contamination in both the star and galaxy samples. Instead, we use the $\rm{SPREAD\_MODEL}$ parameter (from \texttt{SExtractor}) which indicates whether the source is better fit by the PSF model or a more extended model \citep{2012Mohr}. By design, $\rm{SPREAD\_MODEL}$ is close to zero for point sources and positive for extended sources. This estimator has been used in several surveys, e.g., the Blanco Cosmology Survey (BCS)~\citep{2012Desai} and the Dynamical Analysis of Nearby Cluster (DANCe) survey~\citep{2013Bouy}. In particular, we separate stars and galaxies by the following criteria: \begin{equation}\label{eqn::spread}
\begin{aligned}
galaxies: \rm{SPREAD\_MODEL\_I} & > 0.005 \\ \rm{\&} \, \rm{MAG\_i} & < 17.5\\
stars: \rm{SPREAD\_MODEL\_I} & < 0.004,
\end{aligned}
\end{equation}
where $\rm{MAG\_i}$ is the magnitude of the object in $i$ band. This criteria is adapted from~\citet{2018Sevilla-Noarbe}, providing a better separation between stars and galaxies, compared to $\rm{CLASS\_STAR}$, because we take into account the PSF variation in the calculation. The exact values of the thresholds are not extremely important since we will later estimate the photometric redshfits ($z_{phot}$) for each object, as shown in Section~\ref{sec::photoz}. If an object is wrongly identified as a galaxy, we will not obtain a good fit for $z_{phot}$ and the object will be removed from the cluster finding algorithm. More details and different tests to quantify the performance of this star-galaxy statistic can be found in~\citet{2018Sevilla-Noarbe}.

\subsection{Photometric Calibration} \label{sec::photocal}
To calibrate the color and the magnitudes of stars and galaxies, we use Stellar Locus Regression~\citep[SLR;][]{2009High}. SLR adjusts the instrumental colors of stars and galaxies and simultaneously solves for all unknown zero-points by matching them to a universal stellar color-color locus and the known 2MASS catalog. The calibration takes into account difference in instrumental response, atmospheric transparency, and galactic extinction. SLR has been used to calibrate photometry for various surveys including SPT follow-up~\citep{2010High} and the Blanco Cosmology Survey~\citep[BCS;][]{2015Bleem}. The specific implementation of the algorithm that we utilize here is described in~\citet{2014Kelly}. 

For each frame, we use the stellar sources identified in Section~\ref{sec::star_gal} as the starting point. We then perform the stellar locus regression, which simultaneously calibrates all optical colors onto the SDSS system. The absolute flux scaling (or zeropoint) is then constrained via the optical-infrared colors from SLR, combined with the 2MASS point source catalog~\citep{2006Skrutskie}.
 
\subsection{Photometric Verification}
\begin{figure}
\begin{center}
\includegraphics[width=0.95\columnwidth]{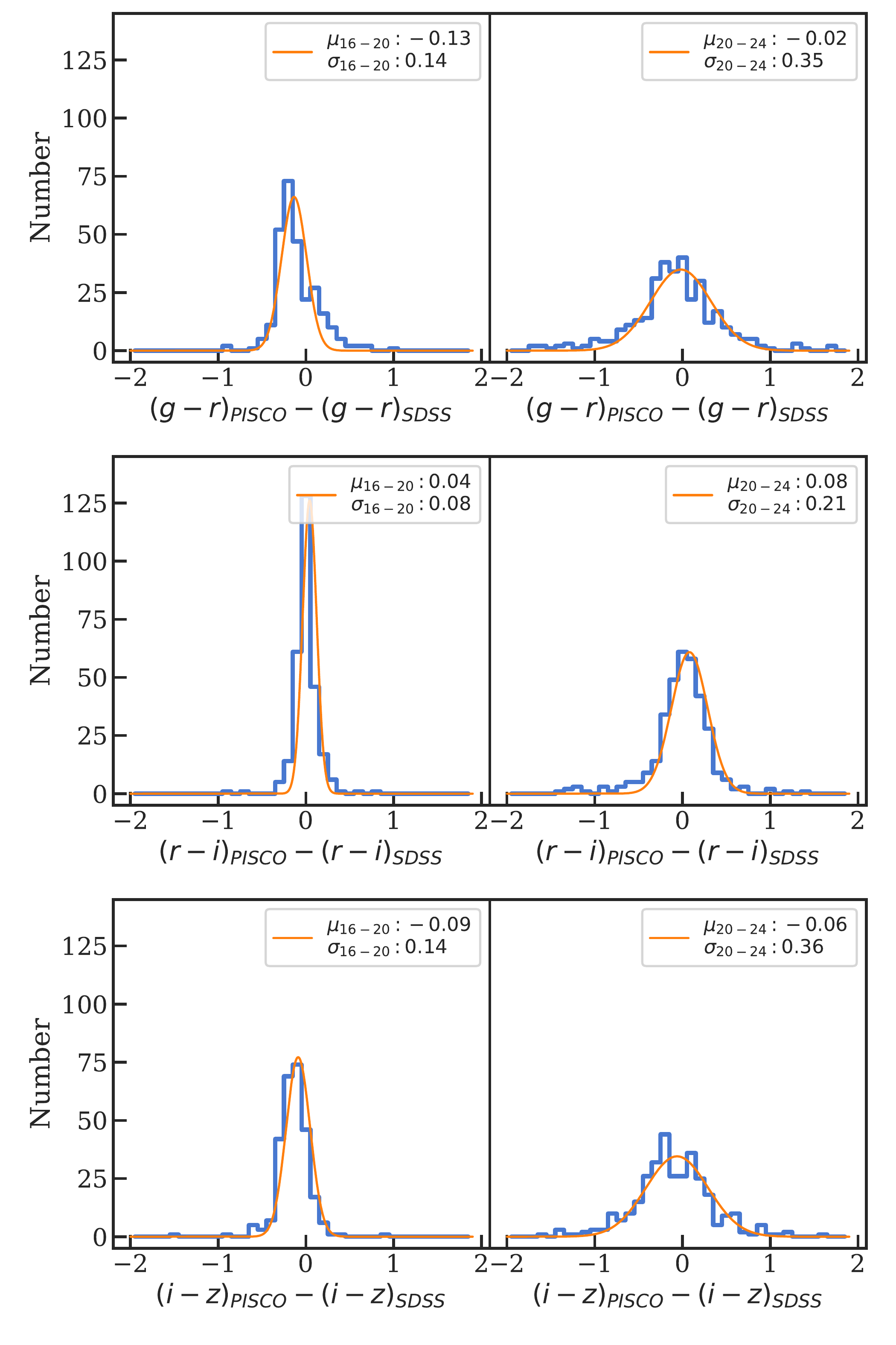}
\caption{The top two panels show a comparison of $g-r$ between PISCO PSF magnitude and SDSS PSF magnitude while the middle two show a comparison of $r-i$ and the bottom two show a comparison of $i-z$. The left panels corresponds to bright objects ($16<i_{\rm{PISCO}}<20$) while the right panels corresponds to fainter ones ($20<i_{\rm{PISCO}}<24$). The orange lines are the gaussian best-fit model with its mean and its standard deviation in the legend. This demonstrate that for bright objects, the scatter of the PISCO calibration from the SDSS is about 0.10-0.14 mag.}
\label{fig::pis_sdss}
\end{center}
\end{figure}

We perform a comparison test to check the accuracy of the photometric calibration. The test is carried out by comparing between the colors ($g-r$, $r-i$, and $i-z$) we obtained from the PISCO pipeline and the SDSS colors. To enable this verification, we observed three fields in our SDSS target list with PISCO, reducing the data using the same PISCO pipeline that we have described above. Galaxies found in the SDSS catalog are matched with objects in our observed PISCO frames based on their celestial coordinates. The objects are plotted in Fig.~\ref{fig::pis_sdss}, showing the offsets between the color from PISCO and SDSS. The scatter ($\sigma$) of the PISCO colors compared to the SDSS colors is around 0.08--0.14 mag for brighter objects ($16<i_{\rm{PISCO}}<20$), which is as accurate as the calibration between SDSS and Pan-STARRS~\citep{2016Chambers}.

\section{Cluster Finding Algorithm} \label{sec::algorithm}
In this section, we describe the new cluster finding algorithm. Because of the nature of our survey, which looks for cluster candidates surrounding X-ray/IR/radio sources, we already have the central location of the cluster, which we assume to be the location of the X-ray/IR/radio source. This means that unlike some optical cluster finding surveys, we do not use a friend-of-friend algorithm~\citep{1982Huchra} to search for the center of the cluster. Instead, we look for an overdensity of galaxies with similar redshifts at the location of the X-ray source. 

Specifically, we search for a peak in the redshift histogram of all the galaxies within the observed fields. Members of a galaxy cluster will have similar redshifts, meaning that finding the peak in the redshift histogram will differentiate between cluster members and field galaxies. The peak location corresponds to the redshift of the galaxy cluster. 

The algorithm is divided into three parts: photometric redshift measurement, aperture selection and background subtraction, and richness correction for high-redshift clusters. 

\subsection{Photometric Redshift}\label{sec::photoz}
The first step of the algorithm is to estimate photometric redshifts of all galaxies in the field. Mid-IR data is included in this step to improve constraints. The sections below describes data acquisition for Mid-IR bands from the Wide-field Infrared Survey Explorer~\citep[\textit{WISE}]{2010Wright}, and the software used for photometric redshift estimates.

\subsubsection{Wide-field Infrared Survey Explorer (\textit{WISE})}
\textit{WISE} is an IR satellite with four IR filters, including $W1$ (3.6 $\mu m$), $W2$ (4.3 $\mu m$), $W3$ (12 $\mu m$), and $W4$ (22 $\mu m$). We select galaxies in the AllWISE Source Catalog, using IRSA's Simple Cone Search (SCS)\footnote{https://irsa.ipac.caltech.edu/docs/vo\_scs.html}, and match them with their optical counterparts from SDSS, Pan-STARRS, or PISCO within a radius of 3". However because the FWHM for $W1$ and $W2$ is rather large ($\sim\!6"$, compared to $\sim\!1"$ for optical data\footnote{http://wise2.ipac.caltech.edu/docs/release/allsky/expsup/\newline sec4\_4c.html}), we cannot separate different optical galaxies from the \textit{WISE} sources, especially at the center of the cluster where large number of objects are presented in a small region. Thus, we only use the \textit{WISE} measurement from both $W1$ and $W2$ as upper limits to help constrain the photometric redshifts.

\subsubsection{Photometric Redshift Estimate}
Each galaxy's photometric redshift ($z_{phot}$) is determined by fitting the photometry in optical and Mid-IR bands to the template spectral energy distribution (SEDs) using the Bayesian Photometric Redshifts (BPZ) code~\citep{2000Benitez,2006Coe}. The BPZ code uses Bayesian inference and priors to estimate photometric redshifts using multi-wavelength broad-band data. We used the default templates, consisting of one early-type, two late-type and one irregular-type templates from~\citet{1980Coleman} and two starburst templates from~\citet{1996Kinney}. We also added WISE filters for $W1$ and $W2$ band. Since there is no response filter for the PISCO optical bands, we convert the photometry from PISCO to SDSS bands and use SDSS response filters instead. Specifically, we convert the photometry to the SDSS system by fitting a linear function of the form: $(g-r)_{SDSS} = A + B(g-r)_{PISCO}$, and likewise for $r-i$ and $i-z$ colors. This amounts to removing the offset shown in Fig.~\ref{fig::pis_sdss}. We apply these corrections to all of the PISCO bands to shift them to the SDSS system. We do not expect the difference in the response filter to have a large impact on the final redshift since PISCO filters are designed to be as similar to the SDSS filters as possible. 

\subsubsection{Redshift Verification}
\begin{figure}[h]
	\begin{center}
		\includegraphics[width=0.94\columnwidth]{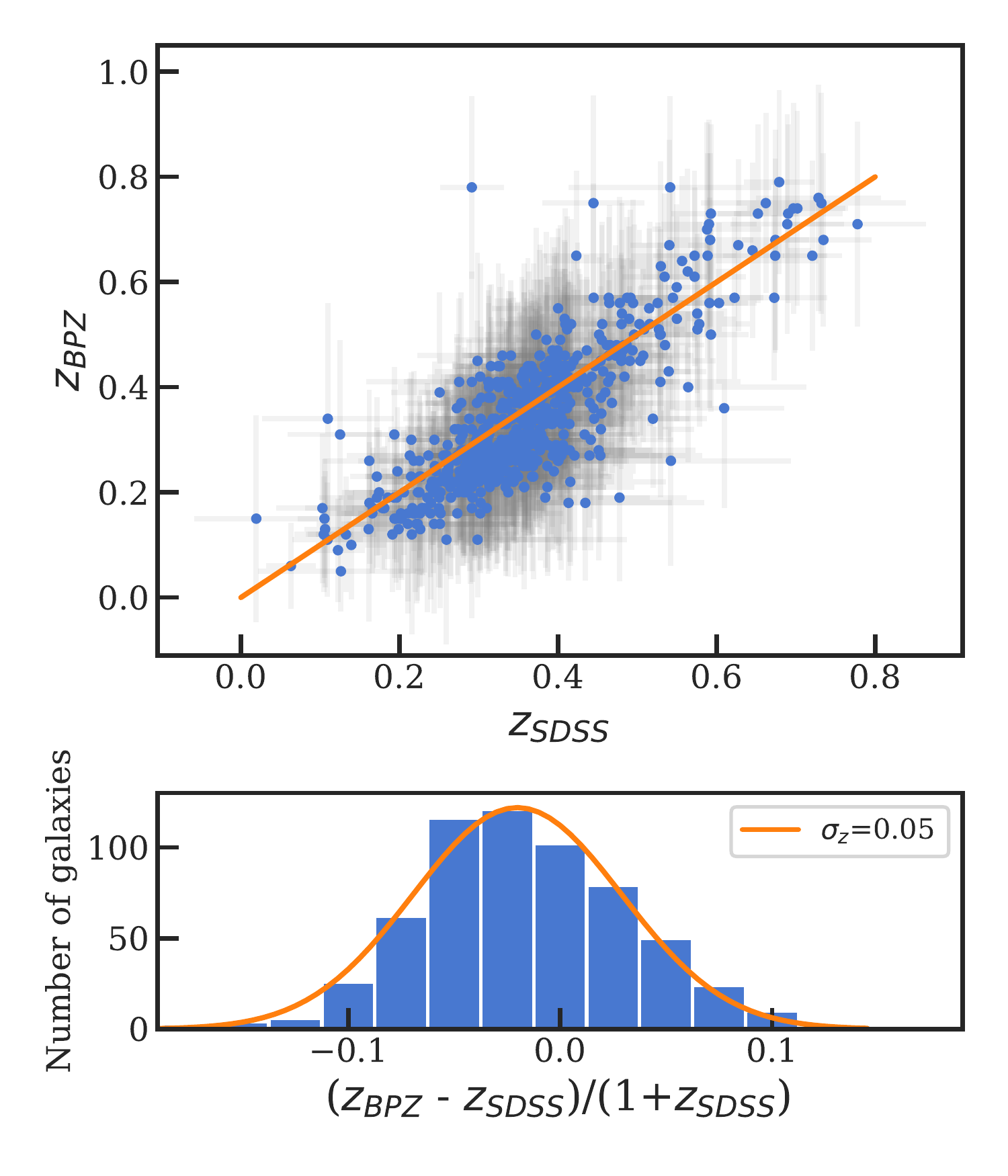}
		\caption{Comparison between the BPZ redshifts and published photometric redshifts from SDSS3. Most of the objects are in an agreement between the two, with a relative scatter of $\sim\!5\%$.}
		\label{fig::sdss-bpz}
	\end{center}
\end{figure}

To verify our photometric redshifts, we compare 612 redshifts from the BPZ algorithm to those publicly available from SDSS3~\citep{2018Abolfathi}. In Fig.~\ref{fig::sdss-bpz}, we show the comparison of BPZ redshifts to those from SDSS3, measuring $\sigma_z/(1+z) \sim\!0.05$. The median offset between BPZ and SDSS3 redshifts is $\sim\!0.03$, which is also less than our typical per-galaxy photometric redshift uncertainty. Utilizing 6 previously-known clusters that were identified in our sample (see Section~\ref{sec::known}), we also find a scatter between our cluster redshifts and the published values of $\sigma_z/(1+z) \sim\!0.012$. Given this overall agreement, we proceed with BPZ redshifts for the full sample.

\subsection{Aperture Selection and Redshift Histogram}
In terms of aperture selection, we choose a simple top hat model with a radius of one arcminute. This allows us to do a more simple correction for the richness value, as discussed in Section~\ref{sec::rich-corr}. Next, we create a histogram representing the redshift distribution of all the galaxies in the selected aperture. Since the peak value includes the background level of field galaxies, we estimate the background distribution by making a redshift histogram of field galaxies in all images of each instrument (SDSS, Pan-STARRS, and PISCO). There are $\sim\!22,000$ background galaxies for PAN-STARRS and PISCO, and $\sim\!27,000$ galaxies for SDSS. We normalized the background histogram for each observation by scaling the total number of objects in the background histogram to be the same as the histogram of interest and subtract from it. 

\begin{figure*}[!ht]
	\begin{center}
		\includegraphics[width=1.98\columnwidth]{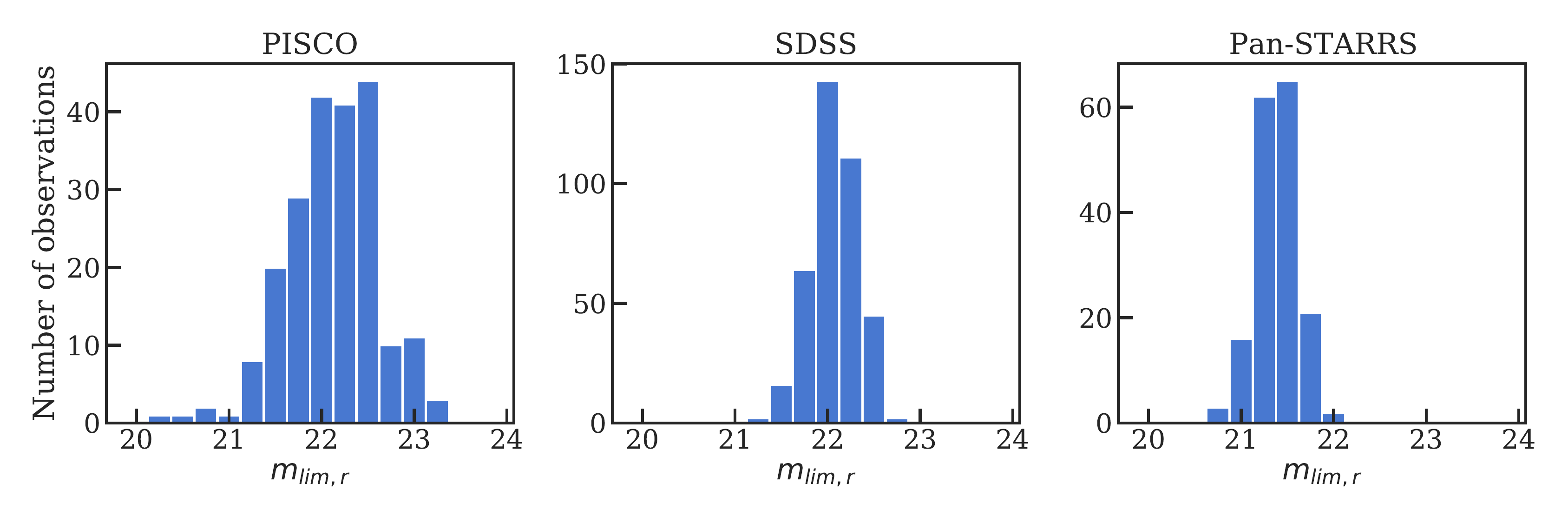}
		\caption{Limiting magnitude distributions for the three telescopes (PISCO, SDSS, and Pan-STARRS, respectively). This figure shows that the magnitude limit for each telescope is fairly uniform. PISCO has significantly more scatter, compared to SDSS and Pan-STARRS which are wide-field surveys.}
		\label{fig::mag_lim}
	\end{center}
\end{figure*}

Fig.~\ref{fig::mag_lim} shows the limiting magnitude for each observation, which demonstrates that the depth of each survey is quite uniform. The detailed calculation for the limiting magnitude is presented in Section~\ref{sec::rich-corr}. For PISCO, we made sure that every observation has the same exposure time of 5 minutes. This uniformity allows us to construct the background for each instrument without large variations in limiting magnitude.

\begin{figure}[!ht]
	\begin{center}
		\includegraphics[width=0.9\columnwidth]{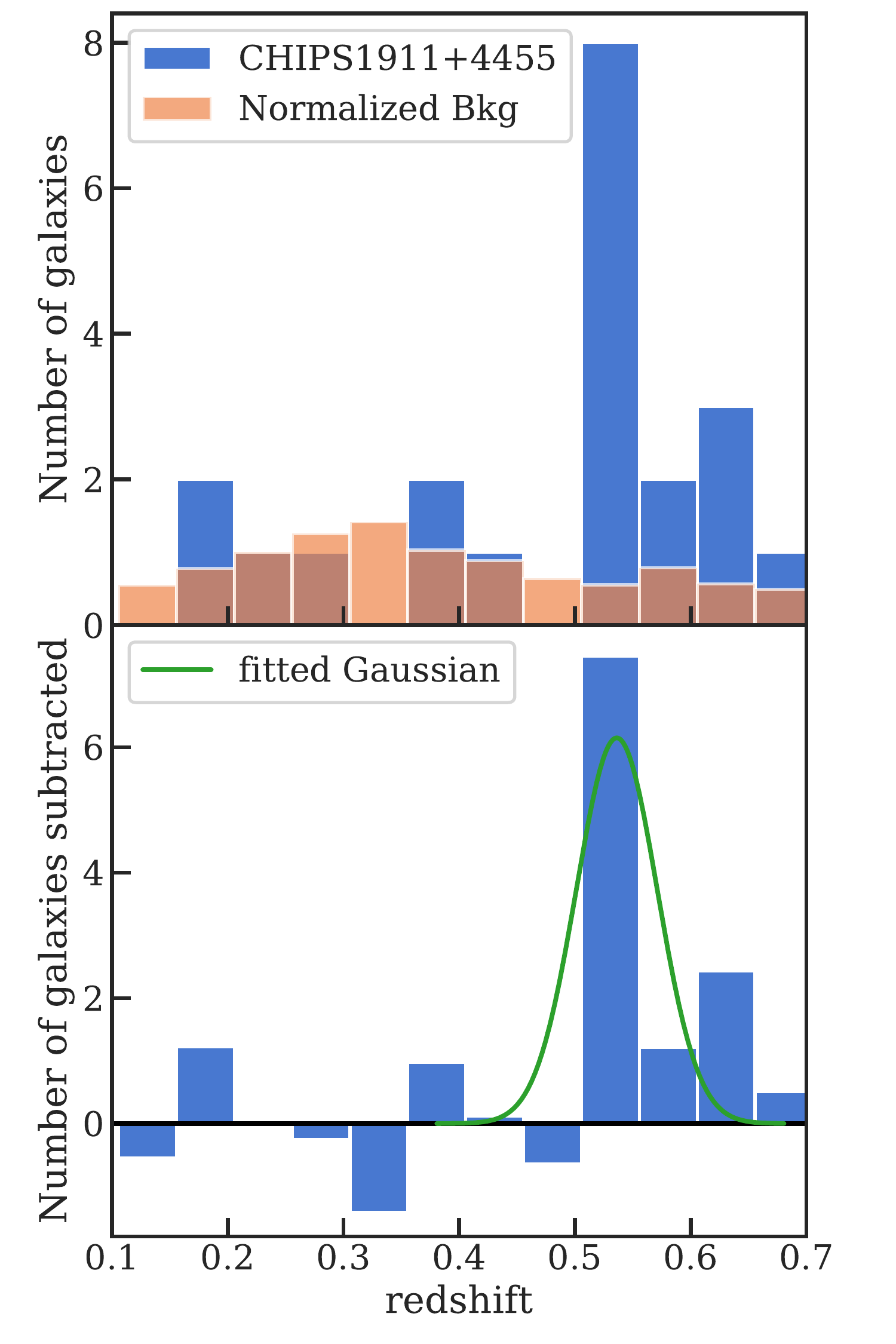}
		\caption{Top: The blue histogram shows the redshift distribution of all the galaxies within one arcmin of CHIPS1911+4455. The orange histogram shows the the renormalized distribution of background galaxies. Bottom: The background-subtracted histogram is shown in blue. The green solid line shows the fixed-width Gaussian fit between the peak and the two neighboring bins to estimate the richness. The significance of the redshift peak at $z=0.53$ strongly suggests the presence of a galaxy cluster.}
		\label{fig::histogram}
	\end{center}
\end{figure}

The top panel of Fig.~\ref{fig::histogram} shows both the redshift distribution of all the galaxies (in blue) and the normalized distribution of background galaxies (in orange). The background-subtracted histogram is then used to search for a redshift peak, as shown in the bottom panel of Fig.~\ref{fig::histogram}. Since the redshifts estimated from the BPZ code have some uncertainty, we fit a fixed-width Gaussian to the peak and the two neighboring bins to get an estimate for the richness (the amplitude of the Gaussian) and the final redshift (the location of the Gaussian). 

\subsection{Richness Correction}\label{sec::rich-corr}
Because observations were made from different optical telescopes and galaxy clusters are located at different redshifts, a correction to the measured richness is necessary to have a uniform proxy for cluster mass across all fields and all redshifts. The two effects we have taken into account include the luminosity function of galaxies and the evolving angular size of galaxy clusters on the sky. We check both effects and find that the luminosity function correction is larger than the angular diameter correction by a factor of $\sim$50--1000, depending on the redshift, so we only consider the luminosity correction. 

Extremely bright objects tend to be rare, compared to fainter objects, implying that cluster candidates at higher redshift will have fewer observable members since the majority of them will be too faint to detect with our current instruments. This correction is used to remedy the galaxy counts to account for galaxies that are below detection limits. The luminosity function ($L(M)$) we used comes from~\citet{2015Wen} which combines the Schechter function $(\phi_s(M))$~\citep{1976Schechter} and the composite luminosity function of the BCGs $(\phi_g(M))$:
\begin{align*}
L(M) = & \phi_s(M)+\phi_g(M) \\
= & \: 0.4\ln{(10)}\phi_*10^{-0.4(M-M_*)(\alpha+1)}\\
& \exp{[-10^{-0.4(M-M_*)}]} \\
& +\frac{\phi_0}{\sqrt{2\pi\sigma}}\exp{\left[-\frac{(M-M_0)^2}{2\sigma^2}\right]}dM,
\end{align*}
where $\alpha$ is the faint-end slope, $M_*$ and $M_0$ are the characteristic absolute magnitudes, $\phi_*$ and $\phi_0$ are the normalization factors. Another effect related to the luminosity function comes from variability in the depth of the survey in different field/telescopes. Specifically, SDSS is deeper (fainter limiting magnitude) than Pan-STARRS. Whereas, PISCO has a large variation within itself, which comes from the variation in the weather conditions when we observed these objects. We estimate the limiting magnitude for each observation by fitting two gaussians to the brightness distribution and using the location of the fainter peak to represent the limiting magnitude. The purpose of the double gaussian fit is to capture the skewness in the brightness distribution, which varies from field to field.

\begin{figure}[!ht]
\begin{center}
\includegraphics[width=0.9\columnwidth]{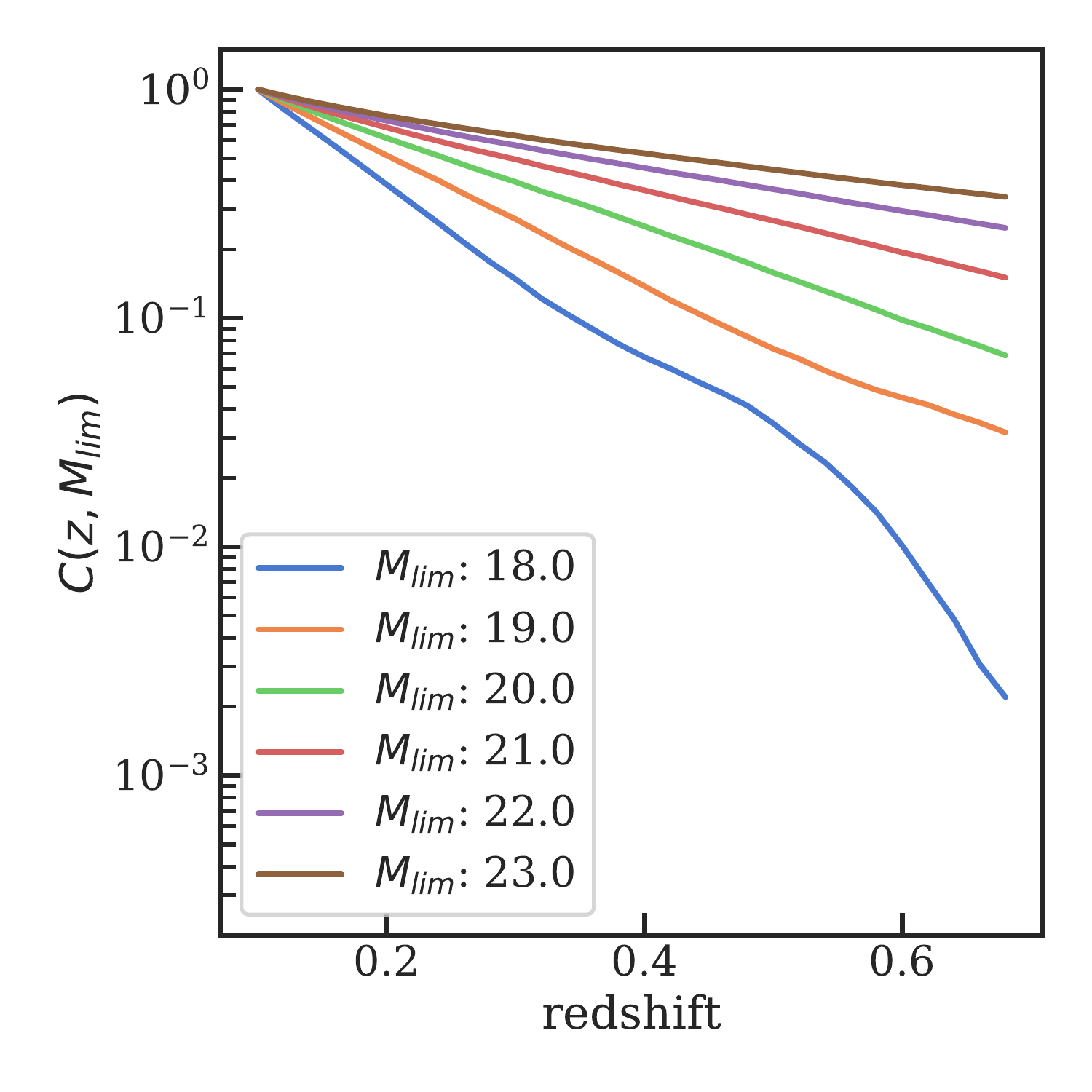}
\caption{Richness correction as a function of redshift and limiting magnitude ($M_{lim}$) due to the luminosity function of galaxies where there are fewer bright massive galaxies, compared to faint smaller ones. Specifically, $R_{true}= \frac{R_{obs}}{C(z,M_{lim})}$, where $R_{true/obs}$ is the corrected and measured richness, and $C(z,M_{lim})$ is the richness correction.}
\label{fig::richness}
\end{center}
\end{figure}

Fig.~\ref{fig::richness} illustrates the richness correction at different redshifts and limiting magnitude ($M_{lim}$). The correction is strongest when we consider high redshift objects with bright limiting magnitudes. The richness is calculated using \begin{equation}\label{eqn::richness}
Richness = N_{obs}\frac{\int^{m_{deep}}_{-\infty} L(M)dM}{\int^{m_{lim}}_{-\infty} L(M)dM},
\end{equation}
where $N_{obs}$ is the number of galaxies found from the survey, $m_{deep}$ is the limiting magnitude of our deepest field ($z=0.1$ at a limiting magnitude of 23), and $m_{lim}$ is the magnitude limit of each field. The typical correction is on the order from 1 to 5.

\subsection{Flux-Limited Nature of Previous Surveys}\label{sec::flux-limit}
\begin{figure*}[!ht]
\begin{center}
	\includegraphics[width=0.65\columnwidth]{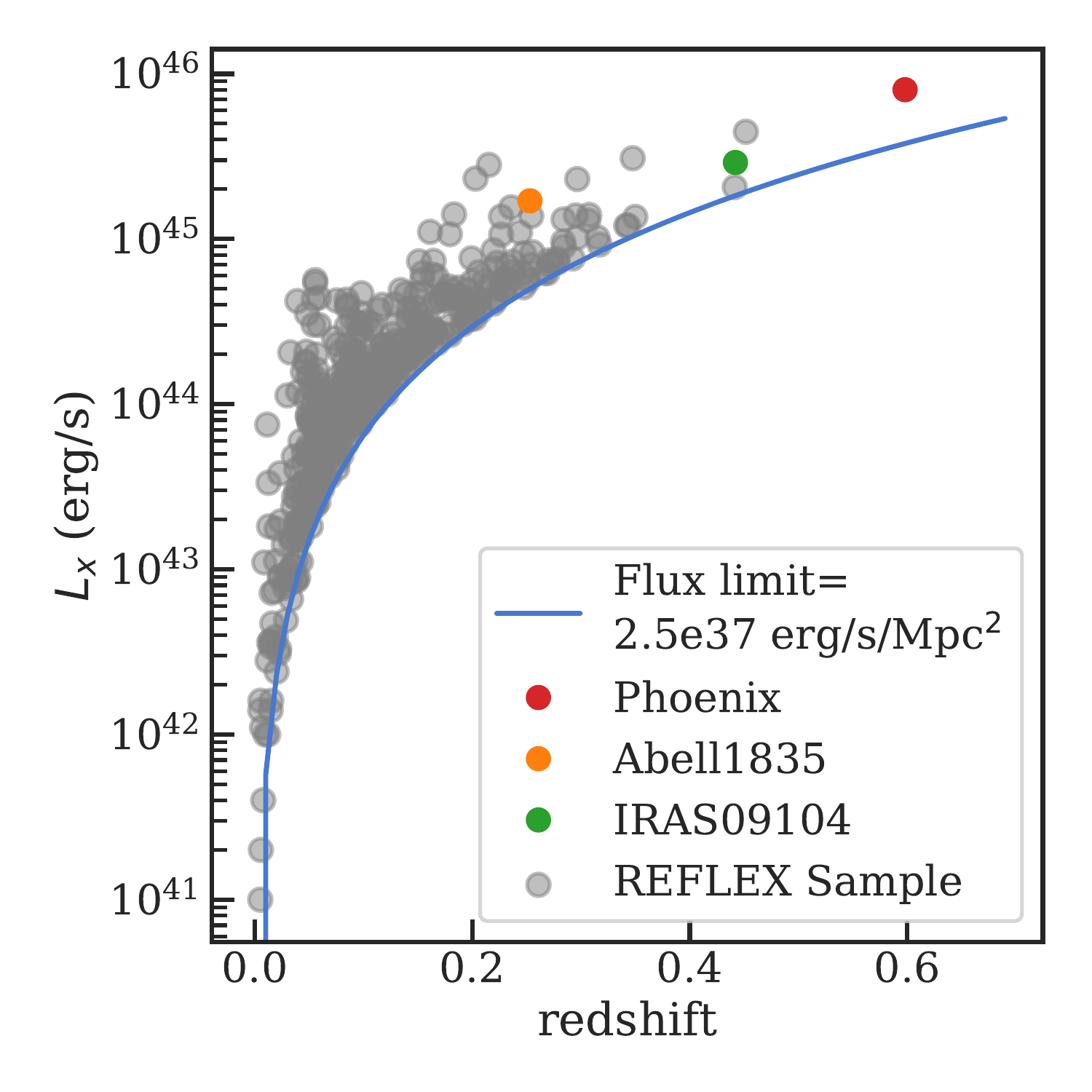}
	\includegraphics[width=0.65\columnwidth]{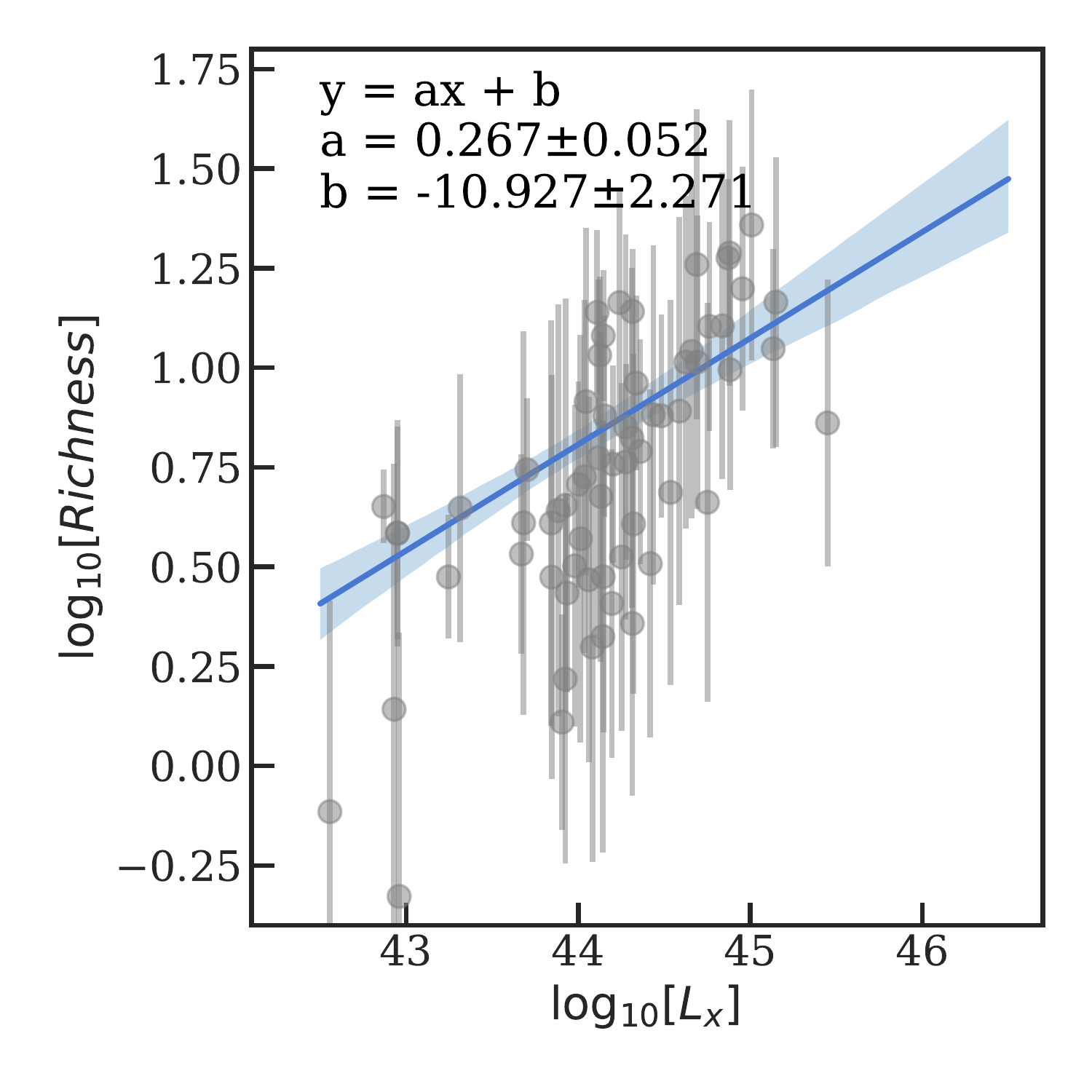}
	\includegraphics[width=0.65\columnwidth]{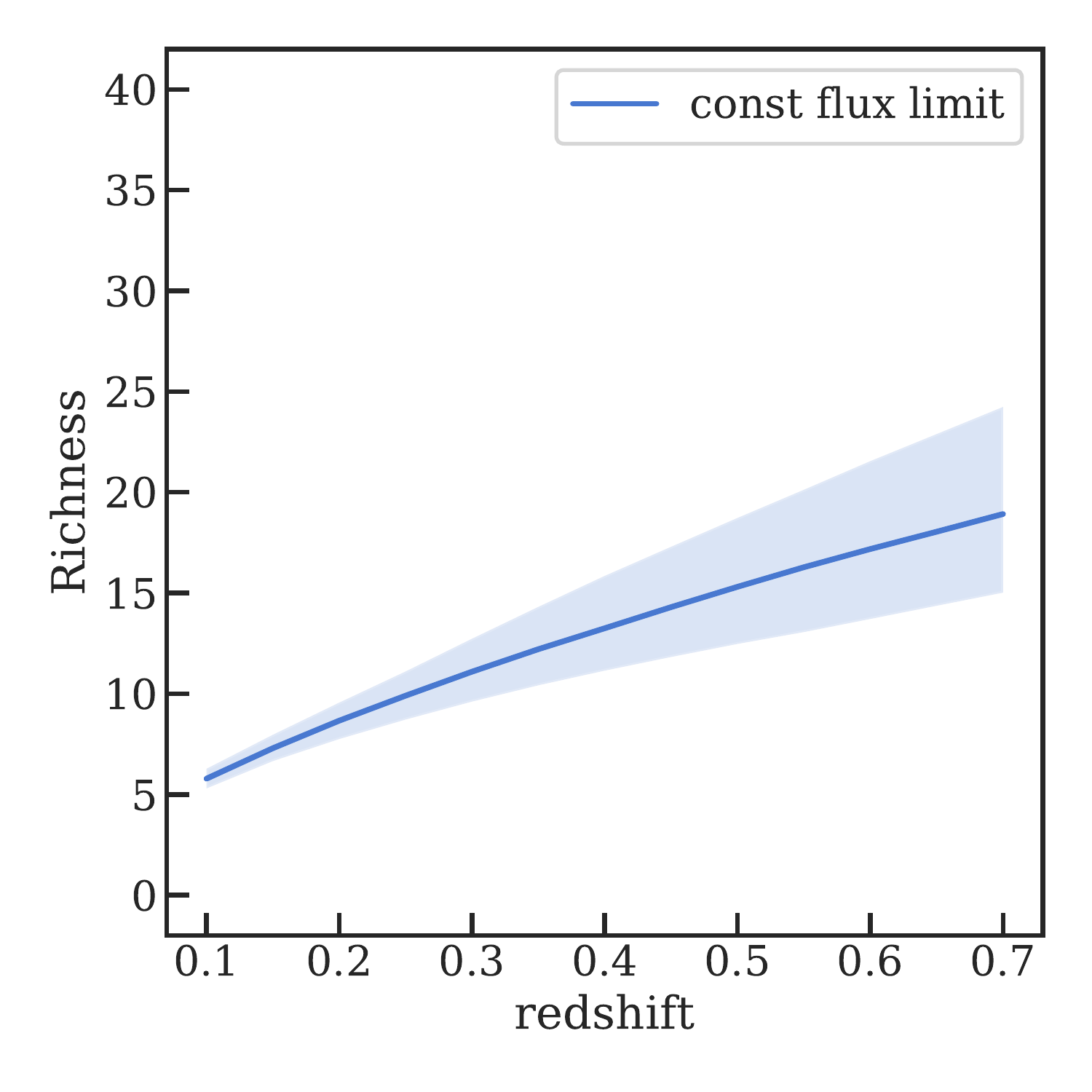}
	\caption{Left: the luminosity vs redshift plot for all of the REFLEX clusters with SDSS data. The plot demonstrates the flux-limited nature of the REFLEX survey. The blue line shows the flux limit = $2.5\times10^{37} \rm{erg/s/Mpc^2}.$ The colored dots show three well-known clusters that should have been detected with the ROSAT catalog. Middle: This plot shows a powerlaw relationship between the luminosity from REFLEX clusters and the measured richness from this work. Right: Richness vs redshift plot with constant flux limit. Assuming the luminosity-richness relation from the middle panel, X-ray-selected clusters from the REFLEX catalog should have richnesses that lie above this line.}
	\label{fig::reflex}
\end{center}
\end{figure*}

The \textit{CHiPS} survey is designed to look for misidentified galaxy clusters in surveys based on data from the ROSAT telescope. One such survey, the ROSAT-ESO Flux-Limited X-ray (REFLEX) Galaxy Cluster Survey~\citep{2004Bohringer}, contains 447 galaxy clusters above an X-ray flux of $\sim\!3\times10^{37}\,\rm{erg\,s^{-1}\,Mpc^{-2}}$ (0.1-2.4 keV) which are all spectroscopically confirmed. The left panel of Fig.~\ref{fig::reflex} shows all 447 clusters in the REFLEX sample on an X-ray luminosity ($L_{\rm x}$) vs redshift plot, with the blue line showing the constant flux limit of the REFLEX sample. It is assumed that this survey has found all of the galaxy clusters with luminosities above this limit.

However, some of the X-ray bright point sources detected by ROSAT are believed to be misidentified massive clusters with extreme central galaxies. One of our goals is to find clusters exceeding the REFLEX flux limit but that are classified as point sources. Since obtaining new optical data is more straightforward to obtain compared to X-ray, we convert this REFLEX flux limit to an optical richness limit which can then be used for our richness cut, as described in Section~\ref{sec::result}.

In order to convert this X-ray flux limit to a richness limit, the optical richness of this sample is required. First, we cross-correlate the REFLEX clusters and SDSS survey, finding 82 clusters that are in both. By running the same cluster finding algorithm as described in this section, we estimate the richness of all 82 REFLEX clusters. Since both the richness and X-ray luminosity are correlated with the total mass of the clusters, we fit a straight line to the log-log plot, as shown in the middle panel of Fig.~\ref{fig::reflex}, to find the relation between the flux limit and the richness limit. For the small number of clusters here, and not accounting for selection effects, we measure an intrinsic scatter between richness and X-ray luminosity $\sigma_{ln\lambda|M} = 0.33\pm0.07$, compared to $\sigma_{ln\lambda|M} \sim\!0.46$ from a sample of SDSS redMapper clusters~\citep{Murata2018}. The last panel of Fig.~\ref{fig::reflex} shows the X-ray flux limit projected onto the richness-redshift plot, via the relationship between richness and X-ray luminosity. Assuming that all clusters follow the same richness--luminosity relation, those systems that lie above this line should have been discovered by the REFLEX survey. In reality, there is significant scatter in the richness--luminosity relation, and those clusters that scatter high in richness at low X-ray luminosity would have been rightfully missed (for example, CHIPS2155-3727).

\section{X-ray Data Reduction} \label{sec::xray}
In addition to the optical survey, we performed X-ray follow-up of 9 promising candidates with Chandra. Three of them are shown in this work. Four candidates were followed up in Chandra Cycle 16 based on an earlier version of our selection (Somboonpanyakul et al. 2018), which yielded the re-observation of a lesser-known cluster, two systems that turned out to be isolated point sources, and CHIPS1356-3421. In Cycle 20, we followed up an additional 5 candidates, which yielded the detection of CHIPS1911+4455. Both of these follow-up campaigns were based on preliminary catalogs and, thus, had an inflated false-positive rate. In this section, we describe the X-ray data reduction process to estimate the total mass and luminosity of these clusters. A more detailed analysis with these data is described in~\citep{2018Somboonpanyakul}.

All \textit{CHiPS} candidates were observed with \textit{Chandra} ACIS-I for 30-40 ks each. The data was analyzed with CIAO~\citep{2006Fruscione} version 4.11 and CALDB version 4.8.5, provided by Chandra X-ray Center (CXC). The event data was re-calibrated with VFAINT mode, and point sources, which are not in the center, were excluded with the \textit{wavdetect} function. The image was produced by applying \textit{csmooth}, which adaptively smoothed an image with maximal smoothing scale of 15 pixels and signal-to-noise ratio between 2.5 and 3.5. 

High angular resolution X-ray images can be used to estimate different properties of the cluster, including the mass and total luminosity. We choose $M_{500}$, the total mass within $R_{500}$, the radius within which the average enclosed density is 500 times the critical density ($\rho_c = 3H_0^2/8 \pi G$), to represent the total cluster mass. We use scaling relations from~\citep{2009Vikhlinin} iteratively to estimate $R_{500}$, which is then used to measure $T_{\rm x}$, $M_g$, and $M_{500}$. Specifically to estimate $M_{500}$, we use the scaling relation with $Y_{\rm x} = M_g\times T_{\rm x}$ 
\begin{align*}
M_{500}=&(5.77\pm0.20)\times10^{14}h^{0.5}M_{\odot} \\ &\times\left(\frac{Y_{\rm x}}{3\times10^{14}M_{\odot}}\right)^{0.57\pm0.03}E(z)^{-2/5}.
\end{align*} 
$Y_{\rm x}$ is chosen as a mass proxy because of its low scatter and insensitivity to the dynamical state of the cluster~\citep{2006Kravtsov,2009Marrone}. More details about the method to estimate $M_{500}$ can be found in~\citet{2011Andersson}.

In addition to mass, we measure the total X-ray luminosity of each cluster. We first extract an X-ray spectrum of all emission within R500, centered on the X-ray peak, and then we fit this spectrum using a combination of collisionally-ionized plasma (\verb|APEC|) and Galactic absorption (\verb|PHABS|). This allows us to estimate the unabsorbed X-ray flux, which we then convert to a rest-frame luminosity given the known redshift.

\section{Results} \label{sec::result}
\begin{figure}
	\begin{center}
		\includegraphics[width=1.00\columnwidth]{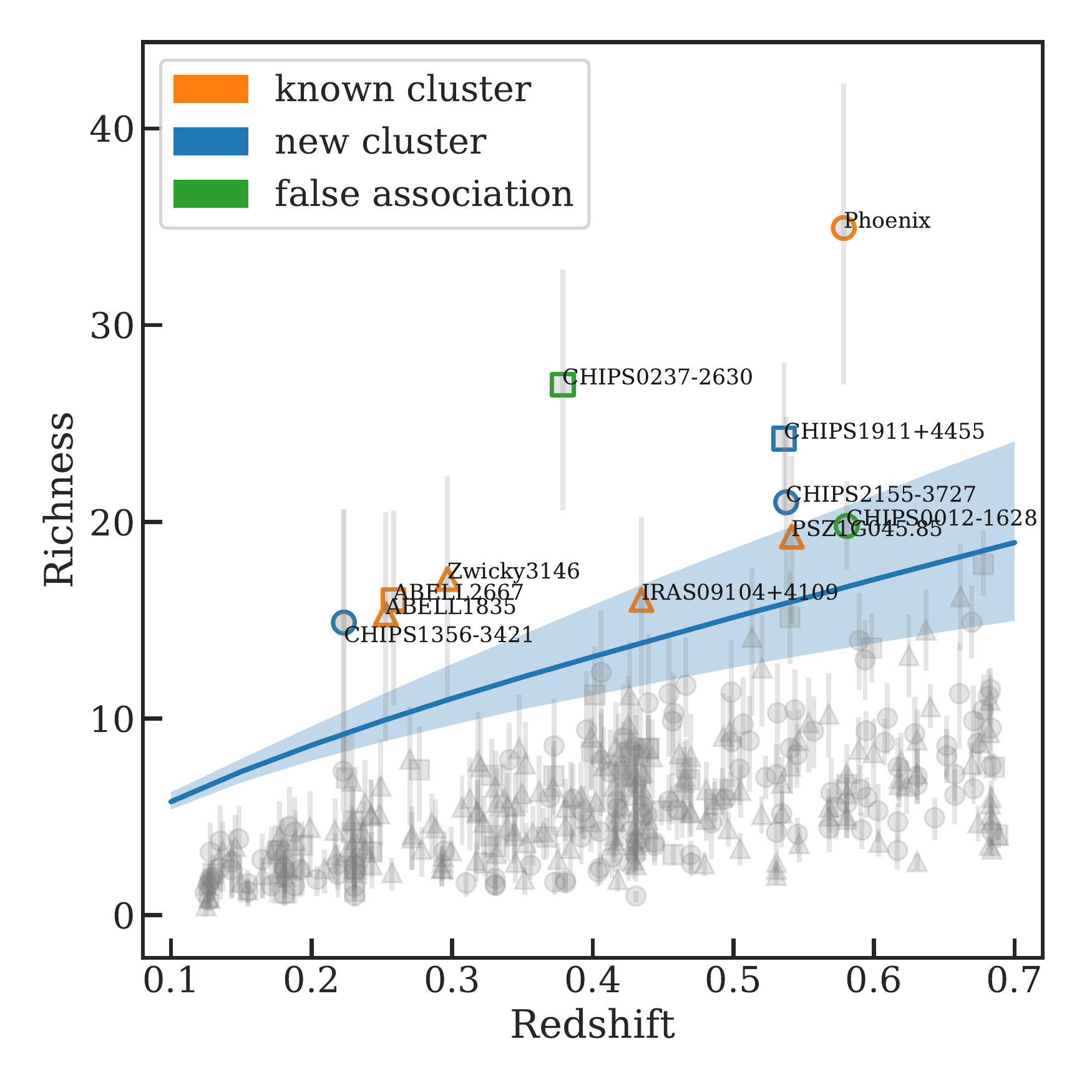}
		\caption{Richness versus redshift for all \textit{CHiPS} candidates. The blue solid line shows the richness cutoff we used for the survey (see Fig.~\ref{fig::reflex}). Each object is indicated with either a circle, a square, or a triangle based on its optical follow-up telescopes, which are the Magellan telescope (PISCO), Pan-STARRS, or SDSS, respectively. The orange color indicates known clusters above the richness limit, the green color indicates a false association, and the blue color indicates a new cluster candidates. All three of the latter systems have recently-obtained \textit{Chandra} X-ray follow-up.}
		\label{fig::all_richness}
	\end{center}
\end{figure}

From Fig.~\ref{fig::all_richness}, we identify 11 cluster candidates by selecting all objects above the solid blue line, which is the richness limit derived in Section~\ref{sec::flux-limit}. The objects below this line are not necessary all isolated AGNs. They may belong to less massive clusters that fall below our selection threshold -- here we only consider very massive clusters that should have been included in surveys such as REFLEX~\citep{2004Bohringer} and MACS~\citep{2001Ebeling}, but were missed due to the presence of an atypical central galaxy.

Using the NASA Extragalactic Database\footnote{https://ned.ipac.caltech.edu} catalog, we search for known clusters within a $3^{\prime}$ radius of the 11 candidates and report, when available, the redshift of known clusters. In Table~\ref{table::richness}, we present all 11 candidates with their celestial coordinates, richness, measured redshifts, instruments used to detect, and known clusters associated with each system.

\begin{deluxetable*}{lrrrrcccc}
	\tabletypesize{\footnotesize}
	\tablecaption{Galaxy cluster candidates/known above the REFLEX flux-limit line in the \textit{CHiPS} Survey\label{table::richness}}
	\tablecolumns{15} 
	\tablehead{\colhead{\textit{CHiPS} Name} & \colhead{RA} & \colhead{DEC} & \colhead{Richness} & \colhead{$z$\tablenotemark{a}} & \colhead{Instruments\tablenotemark{b}} & \colhead{Known Cluster} & \colhead{$z_{\rm NED}$\tablenotemark{c}} & \colhead{Sep (arcmin)} } 
	\startdata
	CHIPS2344-4243 & 356.18375 & -42.72208 & 34.939 & 0.579 & PISCO & Phoenix & 0.596 & 0.388 \\
	CHIPS0237-2630 & 39.365 & -26.5075 & 26.968 & 0.379 & Pan-STARRS & ABELL0368 & 0.22 & 0.395 \\
	CHIPS1911+4455 & 287.75415 & 44.92222 & 24.227 & 0.536 & Pan-STARRS & ... & ... & ... \\
	CHIPS2155-3727 & 328.82791 & -37.46361 & 20.993 & 0.538 & PISCO & ... & ... & ... \\
	CHIPS0012-1628 & 3.145 & -16.46931 & 19.788 & 0.581 & PISCO & ABELL0011 & 0.166 & 3.019 \\
	CHIPS1518+2927 & 229.58292 & 29.45889 & 19.249 & 0.542 & SDSS & PSZ1G045.85 & 0.611 & 0.197 \\
	CHIPS1023+0411 & 155.91374 & 4.18819 & 17.094 & 0.297 & SDSS & Zwicky3146 & 0.2805 & 0.145 \\
	CHIPS2351-2605 & 357.91959 & -26.08403 & 16.027 & 0.258 & Pan-STARRS & ABELL2667 & 0.23 & 0.025 \\
	CHIPS0913+4056 & 138.44167 & 40.93903 & 16.023 & 0.435 & SDSS & IRAS09104+4109 & 0.442 & 0.11 \\
	CHIPS1401+0252 & 210.25876 & 2.88042 & 15.304 & 0.253 & SDSS & ABELL1835 & 0.2532 & 0.055 \\
	CHIPS1356-3421 & 209.023 & -34.3531 & 14.875 & 0.223 & PISCO & ... & ... & ... 
\tablenotetext{a}{These redshifts are photometric redshfits, estimated in Section~\ref{sec::photoz}. We picked a peak of the richness histogram as a cluster redshift. The uncertainty of the redshift is $\sim\!0.025$, which is half a histogram bin width.}
\tablenotetext{b}{PISCO is the imaging instruments on the Magellan telescope in Chile while SDSS and Pan-STARRS are all-sky optical surveys.}
\tablenotetext{c}{Redshift of the known cluster from the NASA Extragalactic Database.}
\end{deluxetable*}

\subsection{Known Clusters Rediscovered} \label{sec::known}
We find 6 of 11 cluster candidates to be well-known clusters via the NED catalog. In general, these clusters can be divided into two classes: starburst-hosting clusters, such as Abell 1835~\citep[$\rm{SFR}\!\sim\!100-180\,M_{\odot}\,yr^{-1}$;][]{2006McNamara} and Zwicky 3146~\citep[$\rm{SFR}\sim\!70\,M_{\odot}\,yr^{-1}$;][]{1994Edge}, or AGN-hosting clusters, such as Abell 2667~\citep{1998Rizza} and IRAS 09104+4109~\citep{1996Crawford}. Another notable example is the Phoenix cluster~\citep{2012McDonald}, which has both a starburst-hosting galaxy and a central AGN. 

The list of ``rediscovered'' clusters include some of the most interesting and well-known galaxy clusters in the nearby Universe. It is an interesting question to ask whether they would have been misidentified by ROSAT as isolated point sources had Abell and Zwicky not performed their optical surveys first. At higher redshift and lower mass, where future surveys like \textit{eROSITA} will probe, this issue will likely be exacerbated, requiring multi-wavelength surveys combining X-ray and, for example, optical or SZ, to fully identify the rich variety of galaxy clusters. 

\subsection{False Associations}
There are two galaxy overdensities that we identify in the background of known clusters (CHIPS0012-1628, CHIPS0237-2630). In both of these cases, the foreground cluster (Abell~11, Abell~368) harbors a BCG that is mid-IR and radio bright, but falls below the richness threshold defined above. At the same time, in both cases, the background overdensity corresponds to a much richer cluster that does \textit{not} harbor an X-ray/IR/radio source at its center. Thus, in these cases, the ``extreme BCG'' is in the foreground, while the massive, ``missed'' cluster is in the background. While these systems are interesting in their own right, for a number of reasons, they do not satisfy the selection requirements of this survey.

\subsection{New Cluster Candidates}
\begin{figure*}[ht]
	\begin{center}
		\includegraphics[width=1.99\columnwidth]{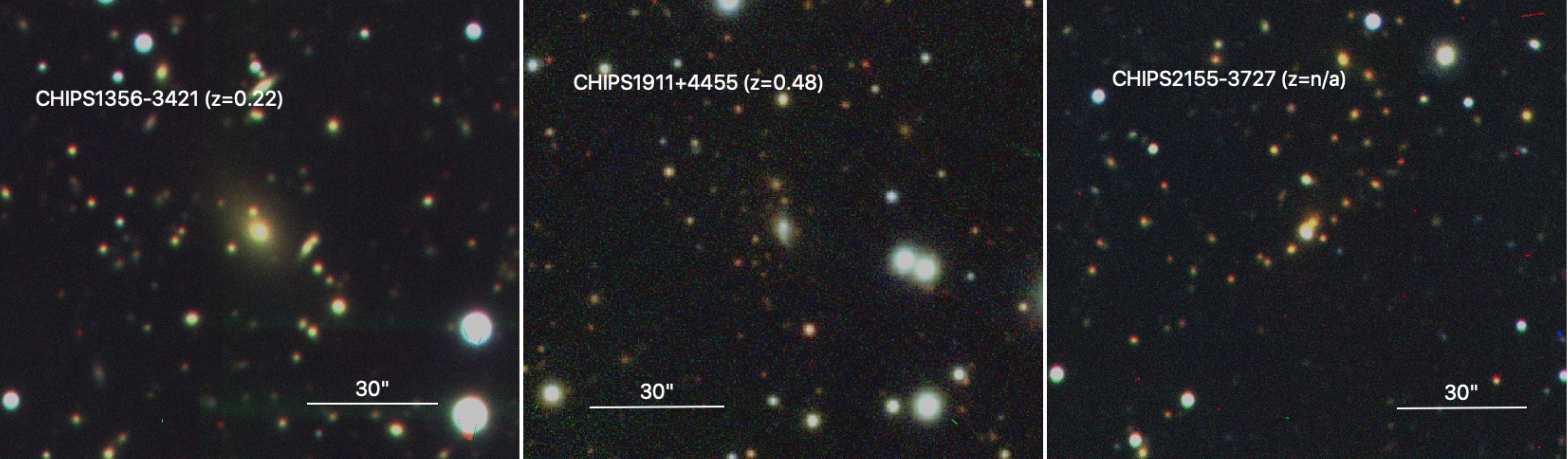}
		\caption{This figure shows $gri$ optical images of all three candidates, including CHIPS1356-3421, CHIPS1911+4455, and CHIPS2155-3727. These new cluster candidates are visually similar in optical, compared to the Phoenix cluster with extremely bright objects in the center. Based on \textit{Chandra} images, CHIPS2155-3727 appears to not be a massive cluster even though the optical image suggests otherwise.}
		\label{fig::optical}
	\end{center}
\end{figure*}

The removal of previously-known clusters and false associations leaves us with a sample of three cluster candidates, all of which are rich enough that they should have been detected by REFLEX or other similar surveys, assuming no scatter in the richness-Lx relation. Fig.~\ref{fig::optical} shows optical images of all three candidates, including CHIPS1356-3421, CHIPS1911+4455, and CHIPS2155-3727. The optical images clearly show an overdensity of red galaxies at the location of the X-ray point source, which is at the center of each field. The three candidates look similar to the Phoenix cluster in that their BCG colors are different from other red member galaxies, implying an active central galaxy.

We followed up all three candidates with new \textit{Chandra} observations over the past two years to search for extended ICM emission, which would confirm the presence of a massive cluster. Even though optical detection of an overdensity of red galaxies alone can often be used to claim discovery of new galaxy clusters, lower-richness candidates can be the result of line of sight alignment from sheets and filaments of galaxies, which can coincidentally increase the numbers of red galaxies on the plane of the sky.

Fig.~\ref{fig::Xray_map} shows adaptively-smoothed \textit{Chandra} X-ray images of all three candidates. The rightmost panel of the figure shows that CHIPS2155-3727 has no (or extremely faint) extended X-ray emission, implying that the overdensity of red galaxies we saw in the optical image in Fig.~\ref{fig::optical} is likely a projection effect. The other two panels show extended emissions around bright point sources in the cores. In Table~\ref{table::chips}, we provide a summary of the X-ray properties ($R_{500}$, $M_{500}$, and $L_{x}$) for the three cluster candidates, derived from the X-ray images. The first two objects are confirmed massive galaxy clusters, with total cluster masses greater than $3\times10^{14}\,M_{\odot}$. With our current dataset, we can only provide upper limits for the mass and total luminosity of a cluster for CHIPS2155-3727. In the follow subsections, we discuss each of these three systems in further detail.

\begin{deluxetable}{lcccc}
	\tabletypesize{\footnotesize}
	\tablecaption{\textit{CHiPS} cluster candidates with \textit{Chandra} follow-up\label{table::chips}}
	\tablecolumns{2} 
	\tablehead{\colhead{\textit{CHiPS} Name} & \colhead{z\tablenotemark{a}} & \colhead{$R_{500}$} & \colhead{$M_{500}$} & \colhead{$L_{x}$\tablenotemark{b}} \\
		& & \colhead{(kpc)} & \colhead{$(10^{14}\,M_{\odot})$} & \colhead{($10^{44}$ erg/s)} } 
	\startdata
	CHIPS1356-3421 & 0.22 & $1300\pm200$ & $6.4\pm3.4$ & $5.9$ \\
	CHIPS1911+4455 & 0.48 & $1075\pm60$ & $6.0\pm0.1$ & $19$ \\
	CHIPS2155-3727 & $\sim\!0.5$ & $<590$ & $<1$ & $<1.1$\\
	\tablenotetext{a}{The redshift is measured spectroscopically for the first two objects, while the third has only a photometric redshift estimate.}
	\tablenotetext{b}{$L_{\rm x}$ is measured from 0.1-2.4 keV with 1 Mpc aperture.}
\end{deluxetable}

\subsubsection{CHIPS1356-3421}
The galaxy cluster surrounding PKS1353-341, also known as CHIPS1356-3421, was the first newly discovered and confirmed galaxy cluster from the \textit{CHiPS} survey with \textit{Chandra} X-ray observations~\citep{2018Somboonpanyakul}. It was missed from other X-ray surveys because of an extremely bright AGN in the central galaxy. Apart from the central QSO, the cluster is an ordinary cool-core cluster with $M_{500}=6.9^{+4.3}_{-2.6}\times10^{14} M_{\odot}$ and $L_{\rm x}=7\times10^{44}\,\rm{erg\,s^{-1}}$ at $z=0.223$. This cluster, the lowest redshift of the three new \textit{CHiPS} clusters, demonstrates how even massive, nearby clusters can be missed if they harbor central X-ray-bright AGN. We measure a star formation rate (SFR), based on archival UV data, in the central galaxy of CHIPS1356-3421, which is roughly a few percent of the cooling rate -- typical of a well-regulated cool core cluster. Given this low star formation rate, we expect that the mid-IR flux is dominated by the central AGN and not by a starburst. More details about this object can be found in our previously published paper~\citep{2018Somboonpanyakul}. 

\subsubsection{CHIPS1911+4455}
CHIPS1911+4455 is the second galaxy cluster we found by the \textit{CHiPS} survey and confirmed with \textit{Chandra} observations. The photometric redshift of the cluster is $z=0.48$. It is our most exciting candidate so far, harboring a very blue galaxy in the center that is surrounded by many red satellite galaxies, similar to the Phoenix cluster~\citep{2012McDonald}. Based on the \textit{Chandra} data we obtained, the total mass and the total size of the cluster are $M_{500}=6.0\pm0.1\times10^{14}\,\rm{M_{\odot}}$ and $R_{500}=1075^{+54}_{-66}$ kpc, respectively, which is as massive as CHIPS1356-3421~\citep{2018Somboonpanyakul}.

With our newly obtained \textit{Chandra} data, we measure the core entropy at $\sim\!10$ kpc to be around 10-20 $\rm{keVcm^2}$, which is as cool as in the Phoenix cluster. However, despite having a blue massive central galaxy and a strong cool core, the system shows a highly-disturbed morphology on both large ($>$100 kpc) and small scales ($\sim\!20$ kpc). Possible scenarios for such a morphology include a recent major merger or a powerful AGN outburst (e.g., ~\citet{2019Calzadilla}). This finding is in opposition to most other known cool core clusters, which are typically highly-relaxed. Additional data will be obtained for this object, including high resolution optical images from the \textit{Hubble} Space Telescope and optical spectra from the Nordic Optical Telescope to look for strong emission lines, a signature of ongoing star formation. A complete analysis of this system is being published in a companion paper, which will include a complete X-ray analysis of the cluster, optical spectroscopy of the central galaxy, and high-resolution \textit{Hubble} imaging of the cluster core.

\subsubsection{CHIPS2155-3727}
\begin{figure*}[ht]
	\begin{center}
		\includegraphics[width=1.98\columnwidth]{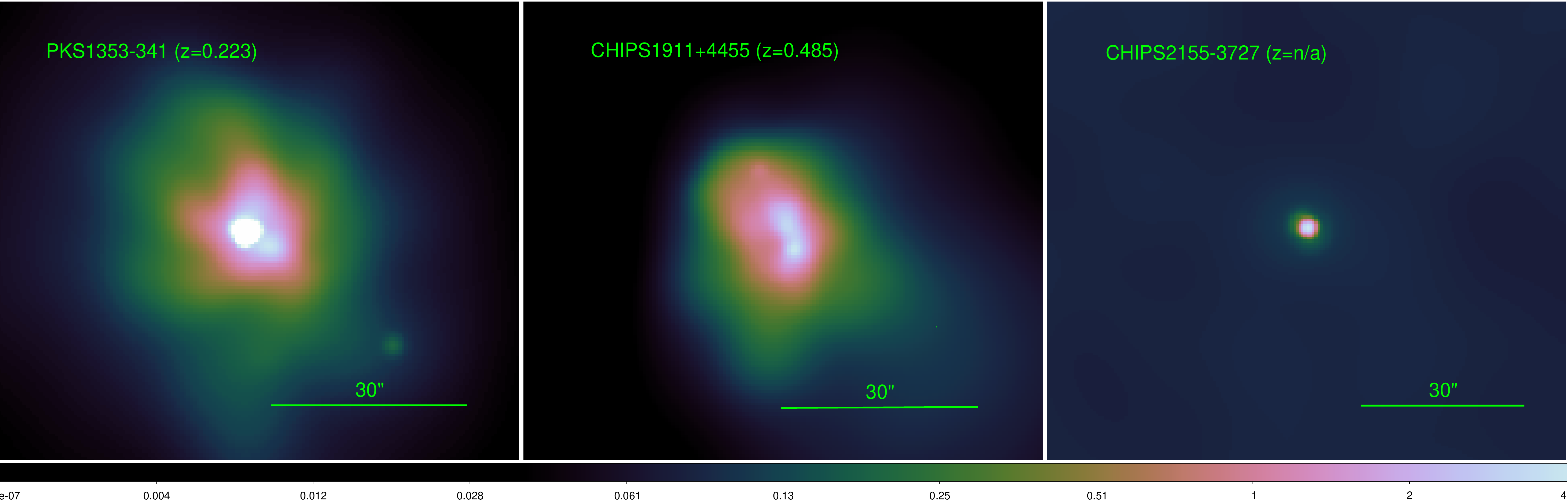}
		\caption{This figure shows X-ray images of the three galaxy clusters candidates: CHIPS1356-3421, CHIPS1911+4455, and CHIPS2155-3727. These \textit{Chandra} images confirm the discovery of two new massive galaxy clusters. These two clusters have a relatively bright core, and CHIPS1911+4455 appears to be a recent merger. The image of CHIPS2155-3727 only shows a bright point source without any extended emission, meaning that it is unlikely to be a massive cluster.}
		\label{fig::Xray_map}
	\end{center}
\end{figure*}

Even though the optical image of CHIPS2155-3727 clearly shows an overdensity of red galaxies at the location of the X-ray source, as shown in Fig.~\ref{fig::optical}, the \textit{Chandra} observation of CHIPS2155-3727 shows no extended emission around the X-ray point source. The non-detection of extended emission in Fig.~\ref{fig::Xray_map} could imply that the overdensity of galaxies is either a projection of a sheet/filament along the line of sight or a smaller galaxy group below our detection threshold. Spectroscopic data is required to determine whether this is simply a projection effect. Based on the X-ray image, the estimated upper limit for the mass of the cluster, if it exists, is less than $1\times10^{14}M_{\odot}$.

\section{Discussion} \label{sec::diss}

\subsection{Updating Flux-Limited Surveys}
\begin{figure}[h]
	\begin{center}
		\includegraphics[width=1\columnwidth]{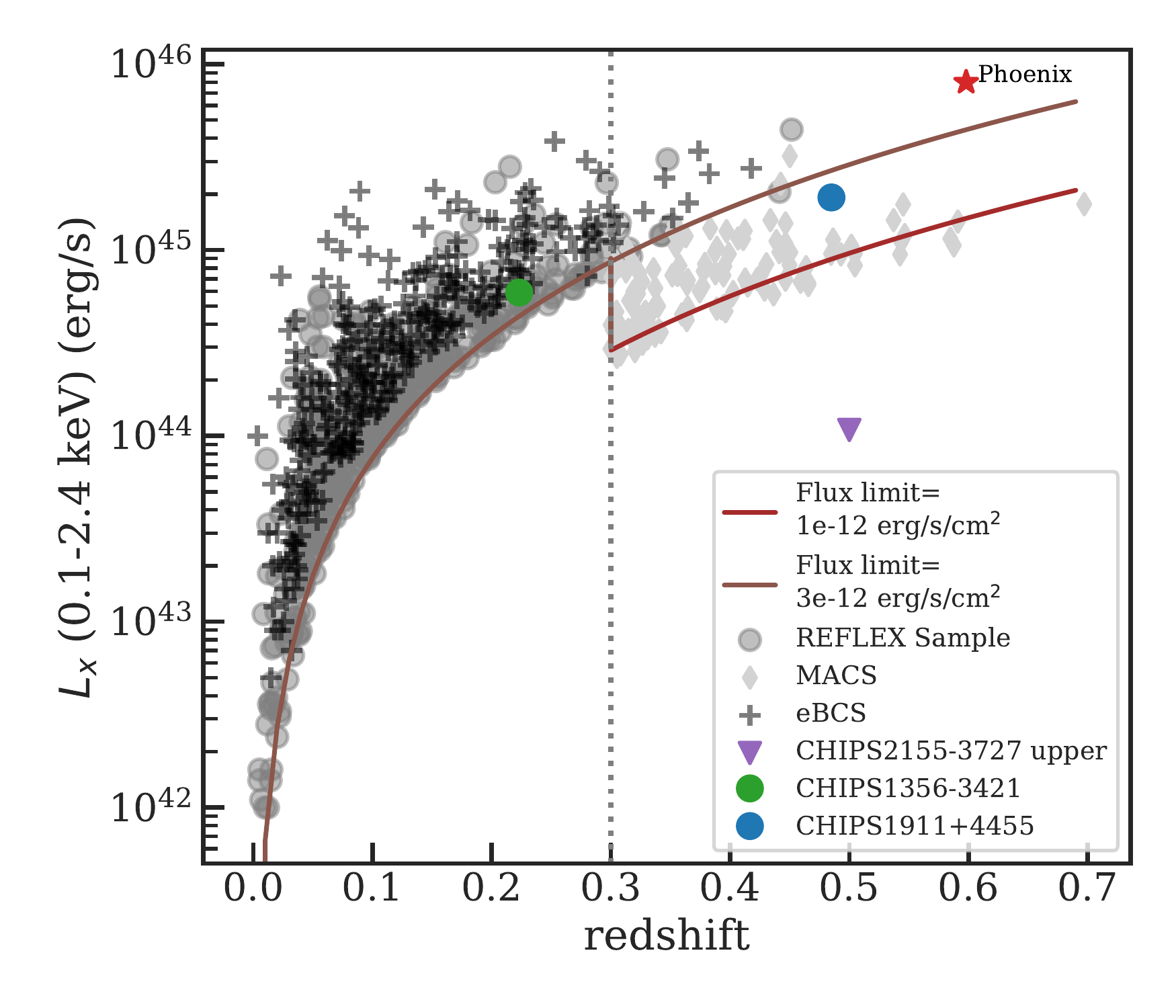}
		\caption{Luminosity versus redshift for clusters from the REFLEX Cluster Survey~\citep{2004Bohringer}, the ROSAT All-Sky Survey Extended Brightest Cluster Sample~\citep[eBCS;][]{2000Ebeling}, and the MAssive Cluster Survey~\citep[MACS;][]{2001Ebeling}. The solid lines show X-ray flux limits, introduced by~\citet{2004Bohringer}, at $1\times10^{-12}\,\rm{erg/s/cm^{2}}$ and $3\times10^{-12}\,\rm{erg/s/cm^{2}}$. This figure shows that the Phoenix cluster, CHIPS1356-3421, and CHIPS1911+4455 should have been identified in the ROSAT data. The purple arrow shows an upper limit for CHIPS2155-3727 which is not detected in our \textit{Chandra} observation.}
		\label{fig::lum}
	\end{center}
\end{figure}

We estimate the rest-frame 0.1-2.4 keV X-ray luminosity, the same as the REFLEX survey, of the new clusters within an aperture of 1 Mpc. Fig.~\ref{fig::lum} shows the X-ray luminosity of newly discovered galaxy clusters and their redshift with respect to clusters from the REFLEX~\citep{2004Bohringer}, eBCS~\citep{2000Ebeling}, and MACS~\citep{2001Ebeling} catalogs. These three cluster catalogs were created by first selecting X-ray bright objects from the ROSAT-All Sky Survey and then confirming via an overdensity of galaxies at a common spectroscopic redshift. The solid lines represent the flux limit of the MACS and REFLEX surveys at $1$ and $3\times10^{-12} \,\rm{erg\,s^{-1}\,cm^{-2}}$, respectively. The fact that clusters from these three catalogs follow closely the aforementioned flux limits highlights the clean selection of X-ray surveys, which are biased towards high-mass systems at high-z, but in a mostly predictable way. However, the figure shows that the two \textit{CHiPS} clusters should have been found by these previous X-ray cluster catalogs, which are all based on the ROSAT data, but were not because of their highly-concentrated X-ray profiles and non-standard BCG colors. This is, similarly, why the Phoenix cluster was not discovered until recently even though it is the most X-ray luminous cluster known~\citep{2012McDonald}. 

\subsection{Rarity of Clusters Hosting Extreme Central Galaxies}\label{sec::rarity}
One of the main goals of the \textit{CHiPS} survey is to find more galaxy clusters with extreme central galaxies (starbursts and/or AGNs) by looking for clusters around X-ray bright point sources which are also bright in the mid-IR and radio. Given that we have only discovered two new Phoenix-like systems, only one of which has an exceptionally-high star formation rate, we can conclude that such rapidly-cooling systems are extremely rare. From this work, we find that the total number of galaxy clusters with extreme central galaxies to be around 10 objects above the ROSAT detection limit. 

To estimate how rare such a system is, we approximate the total population of galaxy clusters found by the ROSAT satellite by combining the REFLEX, eBCS, and MACS samples. Given the total number of clusters detected with the \textit{ROSAT} data to be about 460 in total (0.1$<$z$<$0.7), the occurrence rate of extreme (starbursts and/or rapidly-accreting AGN) central galaxies is $2\pm1\%$. We separate the clusters into two redshift bins to see whether there is any difference. The rate is $2\pm1\%$ for nearby objects ($z=0.1-0.3$) while the rate becomes $5\pm2\%$ for higher redshift objects ($z=0.3-0.7$). At this stage, we do not see any significant difference between the two redshift bins. A deeper and higher resolution X-ray all-sky survey is required to improve our estimate of the occurrence rate of extreme sources in the center of clusters. 

This survey shows that the occurrence rate of clusters hosting extreme central galaxies -- defined as systems with either rapidly accreting supermassive black holes or ongoing, massive starburst -- is extremely low, of order a few percent. This is consistent with the pink/flicker noise temporal statistics observed in the CCA model and related high-resolution hydrodynamical simulations~\citep{2017Gaspari,2019Gaspari}, which predict a 2 dex increase in the SMBH accretion rate $\sim\!1\%$ of the time. Such a rarity of quasar-like blast events is in agreement with the tight, gentle self-regulation driven via CCA (arising from the hot halo condensation), which preserves the cool-core structure for several Gyr. Similar rarity is also consistent with the observed scatter in the SFR at fixed cooling rate~\citep{2016Molendi,2018McDonald}, showing $10\times$ higher SFR in less than $10\%$ of clusters.

\subsubsection{Uniqueness of the Phoenix Cluster}
The Phoenix cluster is one of the most unique clusters found recently~\citep{2012McDonald,2019McDonald}. The central galaxy of the cluster hosts an extremely X-ray luminous AGN with bright radio jets. High resolution optical/X-ray images also reveal a massive cooling flow extended up to hundreds of kpc, which is accompanied by a starburst-hosting BCG. The estimated star formation rate of its BCG is tremendous at $798\pm42\,\rm{M_\odot\,yr^{-1}}$~\citep{2013McDonald}, which is the highest of all known clusters. It seems that the AGN feedback, which has been thought to be responsible for stopping the cooling of new stars in central galaxies~\citep{2012Fabian,2012McNamara}, is not effective in the Phoenix cluster, leading to an extremely high star formation rate in the BCG and the presence of a cooling flow. Nonetheless with only one such system, we cannot fully understand where this system fits in our overall understanding of the cooling/feedback cycle. The \textit{CHiPS} survey was designed in part to find more of these systems.

In this work, we find, at most, one potential analog to the Phoenix cluster: CHIPS1911+4455. This system has a luminous blue galaxy in the center of the cluster, similar to the Phoenix cluster, suggesting the presence of a massive starburst. We will present a detailed analysis of this system, based on ground-based optical spectroscopy and high resolution \textit{Hubble} imaging, in a companion paper. Further, based on in-hand Chandra data, we find evidence that the core may be cooling just as rapidly as in the Phoenix cluster (Somboonpanyakul et al.\ in prep). Considering this, the total number of Phoenix-like clusters are, at most, two (the Phoenix cluster and CHIPS1911+4455) out of $\sim\!460$ systems in a complete X-ray flux-limited sample from the ROSAT-All Sky Survey. This means that the rate of occurrences for such rapidly-cooling systems is less than one percent of the massive cluster population. One explanation for such a rare event is that an intense short-lived cooling of the intracluster medium or a short-lived brightening of the central AGN are a part of the AGN feedback cycle and flickering CCA~\citep{2011Gaspari,2019Prasad}. We can roughly estimate how short this burst of cooling would have been if we find only two such systems at $0.1<\rm{z}<0.7$. Assuming all clusters go through the evolutionary phases in the same manner, within the past $\sim\!5$ Gyr, the rapidly-cooling phase for clusters lasts, on average, for $\sim\!22$ Myr. Since most signatures of star formation last for roughly 20 Myr~\citep{1998Kennicutt}, this implies that almost the full cluster population could go through a short-lived phase of rapid cooling, and we would only expect to observe it (and the subsequent young stellar populations) in a few percent of clusters.

\subsection{Planck Cluster Candidates}
Two of the three clusters, CHIPS1356-3421 and CHIPS1911+4455, have corresponding \textit{Planck} cluster candidates at SNR = 5.76 and 4.64, respectively~\citep{2016Planck-catalog}. Specifically, the \textit{Planck} source at the location of CHIPS1356-3421 is among the 1653 SZ detections in the \textit{Planck} catalog, but it is not a member of the 1203 confirmed detections. \citet{2018Somboonpanyakul} shows that it is in fact a massive cool core cluster at that location. Meanwhile, CHIPS1911+4455 has a weaker signal with SNR=4.64 but has an additional counterpart in an external dataset, specifically a significant galaxy overdensity in the WISE data. These two examples show that we could potentially further utilize the \textit{Planck} catalog of unconfirmed SZ sources to help confirm the existence of these hidden clusters with lower richness than what we are able to achieve currently with the \textit{CHiPS} survey. Even though Planck’s threshold for cluster detection is higher than what we found with the \textit{CHiPS} survey, the Planck catalog also includes many low-significance candidates. Works similar to the \textit{CHiPS} survey will have the potential to help confirm the existence of galaxy clusters in this lower-significance regime since it is highly unlikely that the two completely different techniques will find a cluster candidate at the same location.

\begin{figure*}
	\begin{center}
		\includegraphics[width=1.95\columnwidth]{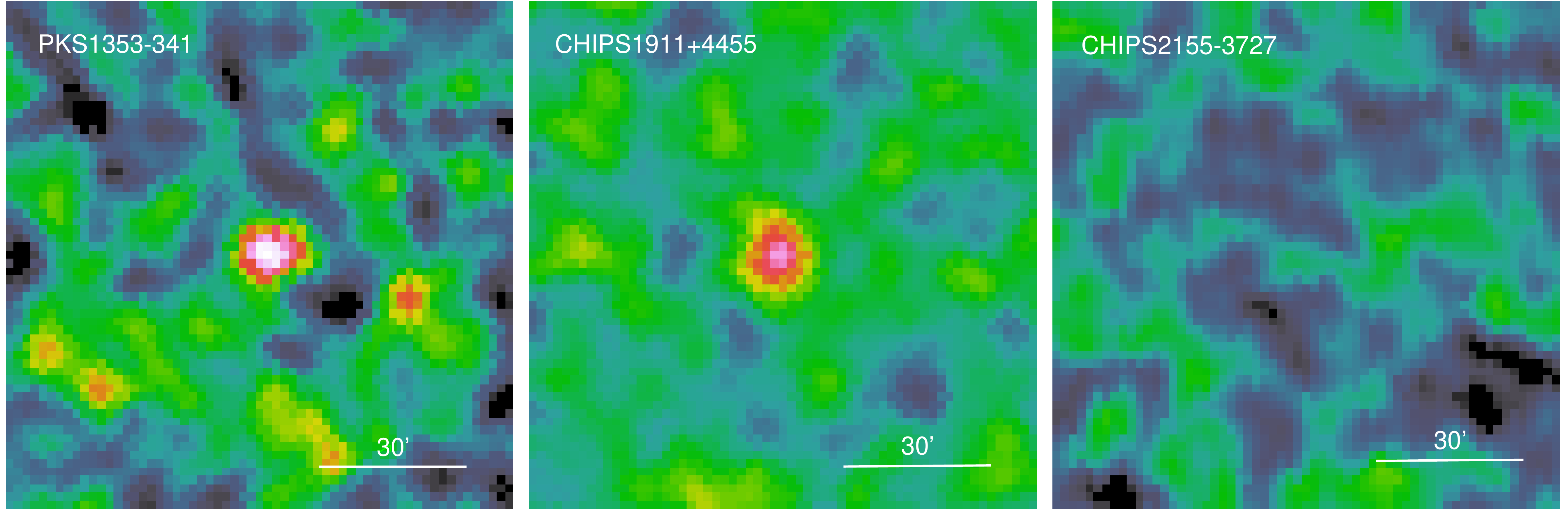}
		\caption{This figure shows $Y$-map images of three galaxy clusters candidates: CHIPS1356-3421, CHIPS1911+4455, and CHIPS2155-3727 (from left to right) from the \textit{Planck} satellite~\citep{2016Planck-catalog}. The two clusters in the left panels are detected with \textit{Planck} and now confirmed with \textit{Chandra}, as described in Section~\ref{sec::result}.}
		\label{fig::sz_map} 
	\end{center}
\end{figure*}

\subsection{Missing known clusters in the survey}

\begin{deluxetable*}{lcccc}
	\tabletypesize{\footnotesize}
	\tablecaption{List of all known clusters with massive star formation rate (SFR$>$60 $\rm{M}_{\odot}\,yr^{-1}$)\label{table::sfr}}
	\tablecolumns{2} 
	\tablehead{\colhead{Cluster Name} & \colhead{z} & \colhead{SFR ($\rm{M}_{\odot}\,yr^{-1}$)\tablenotemark{a}} & \colhead{CHiPS} & \colhead{Reason not found}} 
	\startdata
	RXJ1504.1–0248 & 0.215 & $85\pm9$ & not found & MIR is outside the range \\
	Abell 1835 & 0.2528 & $117\pm24$ & found & \\
	Zw3146 & 0.2966 & $69\pm24$ & found & \\
	IRAS 09104+4109 & 0.4347 & $309\pm120$ & found & \\
	H1821+643 & 0.297 & $447\pm147$ & not found & optical is too bright, not radio bright. \\
	Phoenix & 0.579 & $617\pm200$ & found & \\
	Perseus & 0.0179 & $71\pm20$ & not found & redshift is too low \\
	MACS1931.8–2634 & 0.352 & $263\pm53$ & not found & no match X-ray vs radio \\
	RXJ1532.9+3021 & 0.363 & $98\pm19$ & not found & MIR is outside the range
	\tablenotetext{a}{all SFR numbers come from~\citep{2018McDonald}.}
\end{deluxetable*}

Based on~\citep{2018McDonald}, there are 9 known clusters at $z<1$ that host a massive star-forming galaxy ($\rm{SFR}>60$ $\rm{M}_{\odot}\,yr^{-1}$) in the center. The \textit{CHiPS} survey rediscovered four of them, including Abell 1835, Zwicky 3146, IRAS 09104+4109, and the Phoenix cluster. The other five are the Perseus cluster, H1821+643, MACS1931.8–2634, RXJ1532.9+3021, and RXJ1504.1–0248. In this section, we explain why these five clusters are not detected in the survey. 

The Perseus cluster is the brightest cluster in the X-ray. It is not in our survey because its redshift ($z=0.0179$) falls outside of our interested range of $0.1<z<0.7$. We exclude $z<0.1$ because there are countless optical surveys looking for massive clusters at that redshift range. 

The galaxy cluster surrounding H1821+643 is the only low-redshift ($z=0.299$) galaxy cluster which contains a highly luminous, unobscured quasar in the center~\citep{2010Russell}. However, the quasar H1821+643 is a radio-quiet quasar~\citep{2001Blundell}. It is not included in the \textit{CHiPS} survey because our catalog requires objects to be relatively bright in radio at 1.4 GHz. Furthermore, the \textit{CHiPS} survey is also normalized by the optical images to remove the dependence on redshift; however this also means that we penalize objects which have an extremely bright optical counterpart. These choices were made to reduce the number of candidates to a manageable size for optical follow-up, but will naturally exclude some interesting systems. Thus, we are unable to comment on the occurrence rate of clusters hosting extremely optically-bright or radio-quiet quasars at their center.

MACS1931.8–2634 is another example of a cluster with a powerful AGN outburst amid a major merger event~\citep{2011Ehlert}. However, its X-ray location from RASS-BSC and its radio location from NVSS are 36" apart, which is three times larger than our average distance when matching between the two surveys. 1.4 GHz radio observations from the Very Large Array (VLA) shows a brighter Narrow Angle tail (NAT) radio galaxy 45" to the south of the BCG~\citep{2011Ehlert}. This radio source could be a power jet from a nearby galaxy that is falling into the BCG. With the 45-arcsecond angular resolution of the NVSS catalog, we conclude that the radio location of MACS1931.8–2634 in NVSS is a blended point between the BCG and the radio galaxy. 

The last two clusters, RXJ1504.1–0248 and RXJ1532.9+3021, have relative large star formation rates at $85\pm9\,\rm{M}_{\odot}\,yr^{-1}$ and $98\pm19\,\rm{M}_{\odot}\,yr^{-1}$, respectively. However, they are not within our mid-IR color-cut for our selection which focuses our selections to the Phoenix cluster. In fact, both of them are very close to our selection cutoff from Section~\ref{sec::survey}. This helps to clarify the baseline type of cluster that we expect to find, specifically clusters with SFR $>100\,\rm{M}_{\odot}\,yr^{-1}$ in the BCG.

Both 3C 186 and 3C 254 are also not found in the \textit{CHiPS} survey. This is to be expected since the redshfits for both of them are 1.01 and 0.74, respectively, which is more than our redshift cut at 0.7, as mentioned in Section~\ref{sec::target}.

\subsection{eROSITA}
With the recent launch of the extended ROentgen Survey with an Imaging Telescope Array (\textit{eROSITA})~\citep{2018Predehl} mission in July 2019, an X-ray instrument performing the first imaging all-sky survey in the energy range up to 10 keV, thousands of new galaxy clusters and AGNs will be discovered. 

The \textit{CHiPS} survey helps to predict the potential biases in the \textit{eROSITA} survey, if selection is made based solely on the presence of extended X-ray emission. Some massive groups and clusters with extreme BCGs will appear point-like in the X-ray, similar to what we found with the ROSAT all-sky survey and the \textit{CHiPS} survey. Specifically, the types of system that will be missed include systems where the point source dominates the extended emission (e.g., QSO-central clusters) and systems whose cool core appears point-like (e.g., distant, strongly-cooling systems). With the predicted $10^5$ clusters found with \textit{eROSITA}~\citep{2012Pillepich}, two percent of clusters with extreme central galaxies, as described in Section~\ref{sec::rarity}, is equal to $\sim\!2000$ clusters that \textit{eROSITA} will miss if the survey only characterizes extended X-ray emissions as cluster candidates. \citet{2012Pillepich} also estimated that with \textit{eROSITA} cluster counts and cosmology priors from the \textit{Planck} mission, the uncertainty of $\Omega_{\rm m}$ will be less than two percent. Thus, it is crucial to take into account these missing clusters, which appear point-like.

One proposed solution for \textit{eROSITA} is to allow new X-ray detections be classified as both a point source and an extended source if there is any faint extended emission surrounding a point source. This could potentially help identify even more clusters with extreme central properties. Additionally, the upcoming Vera C. Rubin Observatory, a wide-field telescope with 8.4-meter primary mirror, is expected to be operated by 2021~\citep{2019Ivezic}. The telescope will provide an enormous amount of optical data suitable for following up new cluster candidates. An important note for the optical follow-up is the need to allow the presence of non-typical (i.e., blue) BCGs, which is a current problem in many BCG-identifying codes.

\section{Summary}~\label{sec::summary}
In this work, we present a complete optical description of the Clusters Hiding in Plain Sight (\textit{CHiPS}) survey, a new galaxy cluster survey using both archival (SDSS and Pan-STARRS) and newly acquired data from the Magellan telescope to find new clusters that harbor extreme central galaxies. Our findings are summarized below:

\begin{itemize}
	
	\item By looking at the photometric redshifts of galaxies around X-ray, radio and mid-IR-bright point sources, we have identified 11 cluster candidates. Of these, we rediscovered 6 well-known galaxy clusters with both starburst-hosting and QSO-hosting central galaxies. Two of these candidates were false associations of foreground and background clusters, while the remaining three are previously unknown.

	\item With additional follow-up data from the \textit{Chandra} X-ray telescope for the three new candidates, we confirmed two newly discovered galaxy clusters. We do not detect extended X-ray emission around the other cluster candidate, finding an upper limit on the total mass of $\sim$10$^{14}$ M$_{\odot}$. Details for the first one, CHIPS1356-3421, or the cluster surrounding PKS1353-341, is already published in our pilot paper~\citep{2018Somboonpanyakul}.

	\item We estimate the total mass and the total luminosity of the other new cluster, CHIPS1911+4455. The total mass ($M_{500}$), using the $Y_{\rm x}-M_{500}$ relation, is $6.0\pm1.0\times10^{14}$\,M$_{\odot}$. Whereas, the X-ray luminosity (0.1--2.4 keV) for this cluster is $1.9\times10^{45}\,\rm{erg\,s^{-1}}$. This implies that CHIPS1911+4455 is massive enough to be found by previous X-ray clusters surveys, such as the REFLEX and MACS surveys.
	
	\item We find a massive blue central galaxy in CHIPS1911+4455, pointing to an extreme central galaxy similar to the Phoenix cluster. With the \textit{Chandra} data, we find the core entropy at $\sim$10\,kpc to be as low as in Phoenix, but has a morphology unlike Phoenix and any known strong cool-core cluster. More details about CHIPS1911+4455 will be published in a companion paper.

	\item With the \textit{CHiPS} survey, we find the occurrence rate of clusters that appear as X-ray point sources with bright mid-IR and radio flux to be $2\pm1\%$, and the occurrence rate of clusters with rapidly cooling cores similar to the Phoenix cluster to be less than $1\%$. Such rarity is consistent with the flicker-noise statistics expected during the CCA cycles and with its driven average gentle self-regulation.
	
	\item One of the primary goals of this survey was to determine if the Phoenix cluster is unique. It looks like it is: there is no clusters at $z<0.7$ that have a more massive central starburst within factor of $\sim\!3$ in magnitude. If there was, we would have found it in this survey.
	
\end{itemize}

In general, the discovery of these \textit{CHiPS} clusters emphasize a need for X-ray point source/cluster finding algorithms to allow the possibility of finding both point-like and extended objects at the same time. By limiting the algorithm to only pick out X-ray bright point source, many cluster hosting extreme objects (starbursts/AGNs) were missed in the past. These type of objects are critical in our quest to understand the relation between cooling-flow and feedback from the central BCGs. Lastly, by only finding one new galaxy cluster with a massive starburst galaxy in the center (CHIPS1911+4455), we conclude that the Phoenix cluster is in fact a rare occurrence (less than one percent of all the cluster population). This finding will be important in helping us understand the mechanism of forming a Phoenix-like cluster in the future.

\acknowledgments
Acknowledgments.\\
\textit{Facilities}: Magellan-Clay (PISCO, LDSS-3), SDSS, Pan-STARRS, \textit{Chandra X-ray Observatory} (ACIS) \\
\textit{Software:} astropy~\citep{2018Astropy}, CIAO~\citep{2006Fruscione}, pandas~\citep{2010McKinney}, seaborn~\citep{2016Waskom}, confidence interval calculation~\citep{2011Cameron}

T.\ S.\ and M.\ M.\ acknowledge support from the Kavli Research Investment Fund at MIT, and from NASA through Chandra Award Number GO9-20116X. M.\ G. is supported by the Lyman Spitzer Jr. Fellowship (Princeton University) and by NASA Chandra GO8-19104X/GO9-20114X and HST GO-15890.020-A grants. A.\ A.\ S. and B.\ S. acknowledge support from NSF grant AST-1814719.

The Pan-STARRS1 Surveys (PS1) and the PS1 public science archive have been made possible through contributions by the Institute for Astronomy, the University of Hawaii, the Pan-STARRS Project Office, the Max-Planck Society and its participating institutes, the Max Planck Institute for Astronomy, Heidelberg and the Max Planck Institute for Extraterrestrial Physics, Garching, The Johns Hopkins University, Durham University, the University of Edinburgh, the Queen's University Belfast, the Harvard-Smithsonian Center for Astrophysics, the Las Cumbres Observatory Global Telescope Network Incorporated, the National Central University of Taiwan, the Space Telescope Science Institute, the National Aeronautics and Space Administration under Grant No. NNX08AR22G issued through the Planetary Science Division of the NASA Science Mission Directorate, the National Science Foundation Grant No. AST-1238877, the University of Maryland, Eotvos Lorand University (ELTE), the Los Alamos National Laboratory, and the Gordon and Betty Moore Foundation.

\bibliographystyle{yahapj}
\bibliography{biblio.bib}


\end{document}